\documentclass[journal=jctcce,manuscript=article]{achemso}
\setkeys{acs}{articletitle=true}

\usepackage{xr-hyper}
\usepackage{graphics}
\usepackage{graphicx}
\usepackage{adjustbox}  
\usepackage[dvipsnames]{xcolor}
\usepackage{amsmath}
\usepackage{amssymb}
\usepackage{amsfonts}
\usepackage{gensymb}  %
\usepackage{booktabs}
\usepackage[normalem]{ulem}
\usepackage{hyperref}
\usepackage{placeins}
\usepackage{natbib}

\newcommand{\Molecule}[1]            {\mathit{#1}}
\newcommand{\MoleculePosition}[1]    {\mathbf{R}_{\Molecule{#1}}}
\newcommand{\MoleculeOrientation}[1] {\mathbf{P}_{\Molecule{#1}}}

\newcommand{\NumberOfMolecules}      {N}
\newcommand{\NumberOfConstituent}[1] {\NumberOfMolecules_{\Molecule{#1}}}
\newcommand{\BlueprintCOM}           {\mathbf{R}^{\NumberOfMolecules}}
\newcommand{\BlueprintQuat}          {\mathbf{P}^{\NumberOfMolecules}}
\newcommand{\BlueprintLabframe}      {\mathbf{r}^{\NumberOfAtoms}}
\newcommand{\PositionVector}[1]      {\mathbf{r}_{#1}}

\newcommand{\BlueprintMolecule}[1]   {\mathbf{r}^{\NumberOfConstituent{#1}}}
\newcommand{\BlueprintDFT}[2]        {\mathbf{d}^{#1}(#2)}  %
\newcommand{\BlueprintFit}[2]        {\mathbf{r}^{#1}(#2)}
\newcommand{\BlueprintModel}[1]      {\mathbf{R}^{\NumberOfMolecules}}
\newcommand{\NumberOfAtoms}          {n}
\newcommand{\NumberOfAtomsFromMolecules} {\NumberOfAtoms=\sum\limits_{\mathit{I}=1}^{N}N_{\mathit{I}}}
\newcommand{\MieRep}[1]              {\gamma_{#1}^{(R)}}
\newcommand{\MieAtr}[1]              {\gamma_{#1}^{(A)}}
\newcommand{\UnitCell}               {\mathbb{V}}

\newcommand{\Lattice}                {\mathcal{G}(\BlueprintCOM, \BlueprintQuat, \UnitCell)}
\newcommand{\NumberOfGenomes}        {N_{\rm EA}}

\newcommand{\Population}             {\mathcal{G}^{\NumberOfGenomes}}
\newcommand{\Genome}[1]              {\mathcal{G}_{#1}} %

\title{Reliable computational prediction of supramolecular ordering of complex molecules under electrochemical conditions}

\author{Benedikt Hartl}
\affiliation{Institute for Theoretical Physics and Center for Computational Materials Science (CMS), TU Wien, Wien, Austria}

\author{Shubham Sharma}
\affiliation{FIT Freiburg Centre for Interactive Materials and Bioinspired Technologies, Georges-K\"ohler-Allee 105, 79110 Freiburg, Germany}

\author{Oliver Br\"ugner}
\affiliation{FIT Freiburg Centre for Interactive Materials and Bioinspired Technologies, Georges-K\"ohler-Allee 105, 79110 Freiburg, Germany}

\author{Stijn F. L. Mertens}
\affiliation{Department of Chemistry, Lancaster University, Lancaster LA1 4YB, United Kingdom}
\alsoaffiliation{Institute of Applied Physics, TU Wien, Wien, Austria}

\author{Michael Walter}
\affiliation{FIT Freiburg Centre for Interactive Materials and Bioinspired Technologies, Georges-K\"ohler-Allee 105, 79110 Freiburg, Germany}
\alsoaffiliation{Cluster of Excellence livMatS @ FIT - Freiburg Center for Interactive Materials and Bioinspired Technologies, University of Freiburg, Georges-K\"ohler-Allee 105, D-79110 Freiburg, Germany}
\alsoaffiliation{Frauenhofer IWM, MikroTribologie Centrum $\mu$TC, W\"ohlerstrasse 11, 79108 Freiburg, Germany}

\author{Gerhard Kahl}
\affiliation{Institute for Theoretical Physics and Center for Computational Materials Science (CMS), TU Wien, Wien, Austria}

\begin{document}

\begin{abstract}
We propose a computationally lean, two-stage approach that reliably
predicts self-assembly behavior of complex charged molecules on a
metallic surfaces under electrochemical conditions.  Stage one uses
{\it ab initio} simulations to provide reference data for the energies
(evaluated for archetypical configurations) to fit the parameters of a
conceptually much simpler and computationally less expensive
force field of the molecules: classical, spherical
particles, representing the respective atomic entities; a flat and
perfectly conducting wall represents the metallic surface.  Stage two
feeds the energies that emerge from this force field into
highly efficient and reliable optimization techniques to identify via
energy minimization the ordered ground state configurations of the
molecules.  We demonstrate the power of our approach by successfully
reproducing, on a semi-quantitative level, the intricate
supramolecular ordering observed experimentally for PQP$^+$ and
ClO$_4^-$ molecules at an Au(111)-electrolyte interface, including the
formation of open-porous, self-hosts--guest, and stratified bilayer
phases as a function of the electric field at the solid--liquid
interface. We also discuss the role of the
  perchlorate ions in the self-assembly process, whose positions could
  not be identified in the related experimental investigations.
\end{abstract}

\date{\today}

\maketitle

\section{Introduction}
\label{sec:introduction}

Supramolecular chemistry deals with intermolecular interactions and
structure formation beyond individual molecules, and as such lies at
the basis of many nano- and mesoscopic structures found in biology.
In recent decades, impressive progress in the experimental branches of
this field have resulted in at least two Nobel Prizes in chemistry.
By contrast, the theoretical understanding and especially the {\it in
  silico} prediction of supramolecular ordering has lagged behind
somewhat.  This is easily understood if one considers the sheer size
of the systems under study, requiring in many cases consideration of a
solid substrate, a sufficiently large number of molecular building
blocks or tectons, and a condensed matter medium (i.e. a solvent or
electrolyte solution).  The interaction of these three components,
each with their intrinsic properties, and with optional extrinsic
steering (e.g. by light, heat, electric field), will determine the
observed supramolecular structures and govern the transitions between
them \cite{Cui2018,Mertens2018}.

In this paper, we propose a new theoretical framework to predict
supramolecular ordering of complex molecules at an electrochemical
solid--liquid interface.  The calculations were inspired by recent
experimental work \cite{Cui14acie} in which particularly clear-cut
transitions between supramolecular structures were observed as a
function of the applied electric field at a metal-electrolyte
interface.  The target molecules whose supramolecular ordering is
considered constitute an organic salt that consists of a large,
disc-shaped polyaromatic cation (PQP$^+$) and a much smaller,
inorganic anion (perchlorate, ClO$_4^-$)
\cite{Cui2014ChemCommun,Cui17small}.

The concept of choice to investigate these scenarios would rely (i) on
a faithful description of the properties of the system (notably a
reliable evaluation of its energy) via {\it ab initio} simulations and
(ii) in a subsequent step the identification of the optimized
(ordered) arrangement of the molecules on the substrate by minimizing
this energy via efficient and reliable numerical tools; this
optimization has to be performed in a high dimensional search space,
spanning all possible geometries of the unit cell and all possible
coordinates and orientations of the molecules within that cell. Both
these approaches, considered separately from each other, are
conceptually highly complex and from the numerical point of view very
expensive, which precludes the application of this combined concept
even for a single set of external parameters (such as temperature,
density, and external field); it is thus obvious that systematic
investigations of the self-assembly scenarios of such systems are
definitely out of reach.

In this contribution we propose an approach to overcome these
limitations via the following strategy: in a first step we map the
{\it ab initio} based energies onto the energy of a related classical
model (or classical force field), where the atomistic units
of the molecules are featured as spherical, charged units with
Lennard-Jones type interactions and where the electrolyte is treated
as a homogeneous, dielectric medium; the interaction between the
atomic entities and the metallic surface is modelled by a classical,
perfectly conductive, Lennard-Jones like wall potential.  The as yet
open parameters of the resulting force field (energy- and
length scales, charges, etc.)  are fixed by matching the {\it ab
  initio} energies of the system with the related energies of
this force field: this is achieved by considering
archetypical configurations of the system's building blocks (molecules
and surface) and by systematically varying characteristic distances
between these units over a representative range.  These {\it ab
  initio} energies were then fitted along these `trajectories' by the
parameters of the classical force field: the energy- and
length-scales of the involved interatomic Lennard-Jones or Mie
potentials as well as the atom-wall interaction parameters.

It turns out that this force field is indeed able to
reproduce the {\it ab initio} based energies along these
`trajectories' faithfully and with high accuracy.  Even though the
emerging force field is still quite complex (as it
features both short-range as well as long-range Coulomb interactions
and involves mirror charges) it is now amenable to the aforementioned
optimization techniques which thus brings systematic investigations of
the self-assembly scenarios of these molecules under the variations of
external parameters within reach.

As a benchmark test for our approach we have considered the above
mentioned system, studied in recent experimental investigations: the
cation is PQP$^+$ (9-phenylbenzo[1,2] quinolizino[3, 4, 5, 6-fed]
phenanthridinylium, a disk-shaped polyaromatic molecule), while the
anion is perchlorate, ClO$_4^-$; the self-assembly of these ions on a
Au(111) surface under the influence of an external electric field was
studied.  The high accuracy with which the ensuing energies
calculated from the force field reproduce the {\it ab
  initio} simulation data make us confident about the applicability of
the force field for the subsequent optimization step:
using an optimization technique which is based on ideas of
evolutionary algorithms we have then identified the self-assembly
scenarios of the ions on the Au surface, for a given set of external
parameters (temperature, density, and external
field). These first results provide evidence that our
  approach is quite promising. Tis concept is furthermore completely
  flexible as it can easily be extended to other organic molecules of
  similar (or even higher) complexity.  The computational cost of
this optimization step is still substantial. Therefore we postpone a
detailed, quantitative and, in particular, systematic investigation of
the self-assembly scenarios of the PQP$^+$ and the ClO$_4^-$ ions on
the Au surface for a broad range of external parameters to a later
contribution. Instead we demonstrate in this contribution for selected
sets of parameters that our approach is indeed able to reproduce
several of the experimentally identified self-assembly scenarios.

In this context it has to be emphasized that such a type
of optimization problem is highly non-trivial since the huge number
of possible local minima in the potential energy surface (embedded
in a high-dimensional parameter space) increases exponentially with
the number of particles (and their degrees-of-freedom) of the system
\cite{Stillinger1999}; thus exhaustive search strategies hit the
computational limits or even exceed the capacities of present day
supercomputers.  Yet another complication in structure prediction
algorithms is caused by the fact that different polymorphs of a
system can be kinetically trapped and a vast number of other minima,
having values of the internal energy comparable to the global
minimum may also play an important rule in structure formation
processes \cite{Hofmann2018,Stillinger1999}.

At this point we owe an explanation to the reader why we
have chosen the possibly unconventional approach. Of course, it is
obvious that an optimization of the molecular configurations on the
basis of full {\it ab initio} calculations is from the computational
point of view by far out of reach. However, one can argue that
suitable force fields (available in literature) or machine-learning
(ML) potentials such as high-dimensional neural network potentials
\cite{Blank1995,Lorenz2004,Behler2007,Behler2007a,Behler2016,Schutt2017,Unke2019},
kernel-based ML methods\cite{KernelMethods} (such as Gaussian
approximation
potentials\cite{Gabor2010,Rupp2012,Gabor2013,Gabor2016,Ceriotti2016})
or more specialized, effective potentials for selected molecular
motives such as the SAMPLE
approach\cite{Hofmann2017,Hofmann2018,Hofmann2019} might represent a
more conventional approach to tackle this problem (note that the
field of ML potentials is rapidly growing and the above list is far
from comprehensive). Such arguments represent fully legitimate
objections against our approach.

The problem we are addressing in this contribution is
however a non-standard problem and thus requires to be treated with
a custom force field:  the justification for our strategy is that
we wanted to endow the atomic units of the molecules with ``real''
physical properties (such as ``size'' or ``charge''), which will help
us to perform the second step in our structural search that we have
envisaged (and that we are currently working on): as the computational
costs of our approach are still considerably, large scale investigations
are still prohibitively expensive. In an effort to overcome these
limitations we plan to proceed to even more coarse-grained models
which grasp, nevertheless, the essential features of our complex
molecules. On the basis of such models we would then be able to
identify with rather low computational costs first trends in
structural identification processes. Investigations along this
direction will be explored in forthcoming contributions.

Finally we point out that we are well aware of the
limitations and deficiencies of our present model. Features such as
the response of the metallic electronic distribution of the gold
surface due to the presence of an external bias, variable
electrostatic properties of the molecular species (allowing thus for
polarization effects), or a space dependent permittivity cannot be
included in our concept.  However, at this point it is fair to say
that, to the best of our knowledge, none of the aforementioned
alternative approaches (such as the use of conventional force fields
or machine learning frameworks) are able to take all these effects
faithfully into account, either.

The manuscript is organized as follows: In Section~\ref{sec:system} we
describe the essential features of the experimental setup, introduce
an {\it ab initio} and a classical representation thereof and discuss
the mapping procedure between those different instances.  In
Section~\ref{sec:self_assembly} we put forward the memetic
optimization procedure based on ideas of evolutionary algorithms in
order to identify ordered ground state configurations of complex
molecules under electrochemical conditions and in
Section~\ref{sec:results} we present selected numerical results which
demonstrate a semi-quantitative agreement with the experimentally
observed self-assembly scenarios of PQP$^+$ and ClO$_4^-$ ions on an
Au(111)-electrolyte--interface under the influence of an external
electrostatic field.  We conclude our findings in
Section~\ref{sec:conclusion}.

\section{The system and its representations}
\label{sec:system}

\subsection{The system}
\label{subsec:system}

\begin{figure}[htbp]
\begin{center}
\includegraphics[width=\textwidth, clip]{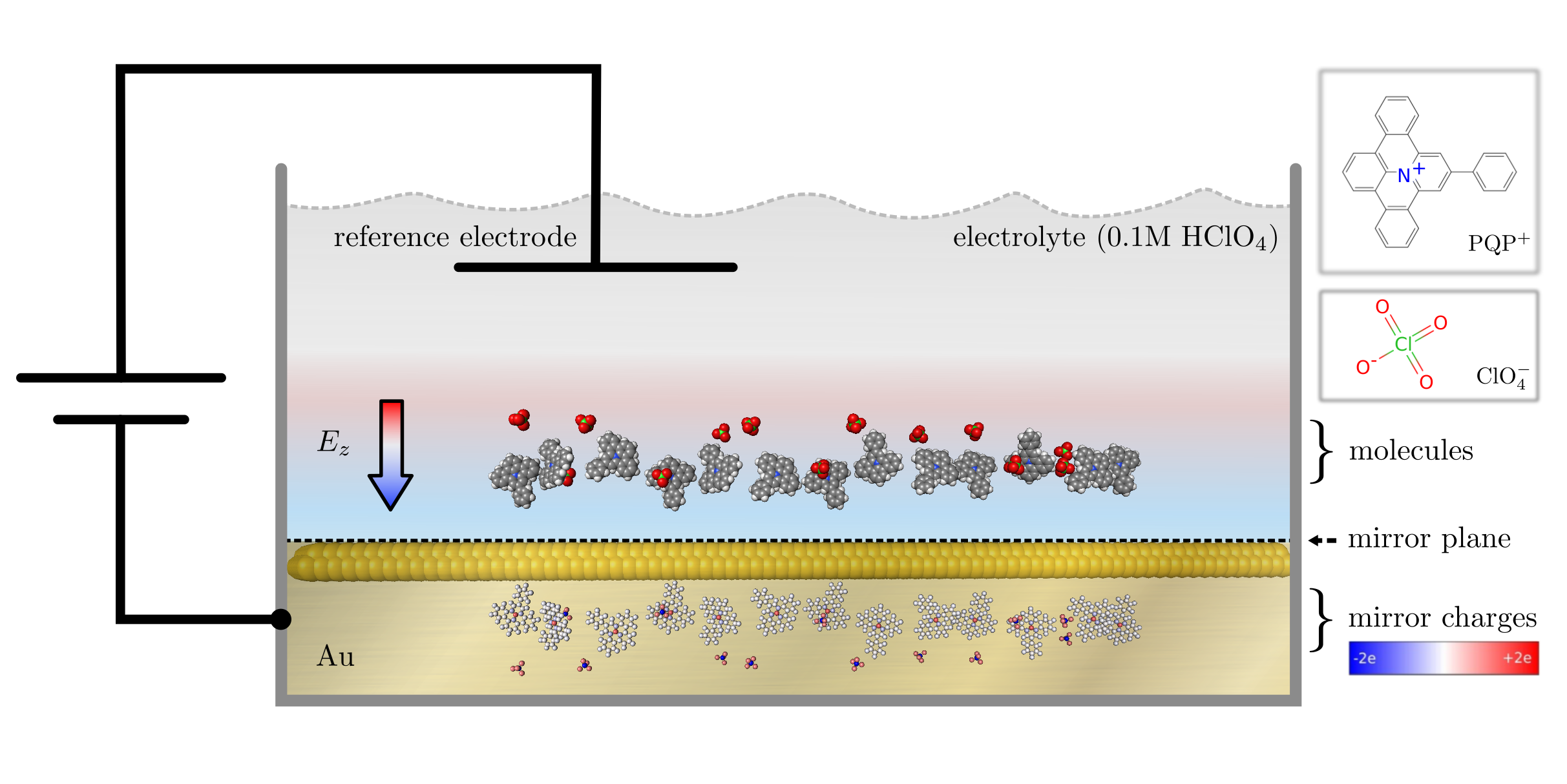}
\caption{(color online) Schematic visualization of the experimental
  setup to control the pattern formation of PQP$^+$ (and ClO$_4^-$)
  molecules (structure formulas given in top right insets) close to a
  Au(111)-surface: two Au-layers are explicitly shown, the golden,
  shiny area represents the conductive Au-bulk, the black dashed line
  marks the surface of the electronic density which we interpret as
  mirror plane.  The ions are immersed into an electrolyte (gray,
  shaded region), which is considered as an effective, homogeneous
  medium.  In the region close to the Au surface (red to blue shaded
  areas) a homogeneous, electrostatic field $E_z$ (bold, colored
  arrow), oriented in the $z$-direction, features the electrostatic
  potential drop between the Au-surface and the reference-electrode
  inside the electrolyte.  The colors of the atoms in the electrolyte
  correspond to their type, while the color of the mirror-atoms
  (located in the Au-bulk) specify their partial charges, quantified
  by the colorbar (see bottom right) in units of the electron charge,
  $e$.  }
\label{fig:experiment:atomistic}
\end{center}
\end{figure}

Both the DFT calculations and the related force field are
based on a framework that mimics the essential features of the
experimental setup, put forward (and discussed) in \cite{Cui14acie};
this framework is schematically depicted in
Fig.~\ref{fig:experiment:atomistic}: PQP$^+$ and ClO$_4^-$ ions are
immersed into an electrolyte (aqueous 0.1M perchloric acid).  From
below, the system is confined by a Au(111) surface, which in the
experiment serves as the solid substrate for adsorption.  An electric
field, E$_z$, can be applied between a reference electrode located
within the electrolyte and the Au surface.  The PQP$^+$ and the
ClO$_4^-$ ions are first treated via DFT based {\it ab initio}
calculations (see Subsection~\ref{subsec:system_ab_initio}).  The
calculated energies are then used to fix the force fields
of classical particles (notably their sizes, energy parameters, and
charges) which represent the atomic entities of the respective ions;
the interaction between the atomic entities and the Au(111) substrate
is described by means of a classical wall-particle force
  field (see Subsection~\ref{subsec:system_classical}).  Throughout
the electrolyte molecules are not considered explicitly.  The
electrolyte is rather assumed to be a homogeneous effective medium
with a permittivity of water, i.e.~$\epsilon_r=78.36$, at
$T=25\celsius$ \cite{Kaatze1989,Hamelin1998,Vargaftik1983},
corresponding to the temperature at which the experiments by Cui {\it
  et al.}  \cite{Cui14acie} were carried out and assuming that the low
concentration of perchloric acid does not change the value of
$\epsilon_r$ substantially
\cite{Hasted1948,Nir2016,Mollerup2015,Fumagalli1339}.  Hence, in this
contribution we use `electrolyte' as synonym for `solvent' unless
explicit use is required.

We emphasize at this point that in the experiment, an
exact specification of the electric field strength is not possible:
as detailed in the Supplementary Information of \citenum{Cui14acie},
the authors of the related experimental investigations have
estimated rather the degree of charge compensation on the Au surface
by the adsorbed PQP$^+$ ions as a function of their changing
coverage, which does not allow to estimate the electric field
directly. This fact limits the degree of quantitative comparison
between experiment and theory.

\subsection{{\it Ab initio} simulations}
\label{subsec:system_ab_initio}

The density functional theory calculations were performed with the
software package GPAW \cite{Mortensen05prb,Enkovaara10jpc} and the
structures handled by the atomic simulation environment
\cite{Larsen17jpc}.  The electronic density and the Kohn-Sham orbitals
were represented within the projector augmented wave method
\cite{Blochl94prb}, where the smooth parts were represented on real
space grids with grid spacing of 0.2 \AA\ for the orbitals and 0.1
\AA\ for the electron density.  The exchange-correlation energy is
approximated as proposed by Perdew, Burke and Ernzerhof (PBE)
\cite{Perdew96prl} and weak interactions missing in the PBE functional
are described as proposed by Tkatchenko and Scheffler (TS09)
\cite{Tkatchenko09prl}.  The TS09 approximation assumes that long
range dispersive contributions are absent in the PBE functional, such
that these can be applied as a correction. The total energy which is
written as

\begin{equation}
  E = E_{\rm PBE} + w_S E_{\rm vdW}
  \label{eq:EDFT}
\end{equation}
where $E_{\rm PBE}$ is the PBE energy and $E_{\rm vdW}$ is the TS09
correction. We have introduced a weight factor $w_S$ that will allow
to incorporate electrolyte effects into the dispersive contributions
as discussed below.  For interactions in vacuum $w_S=1$.  The presence
of the aqueous environment on the electronic and nuclear degrees of
freedom included in $E_{\rm PBE}$ is modeled by a continuum solvent
model \cite{Held14jcp}.

Molecular interactions are studied on simulation grids with Dirichlet
(zero) boundary conditions.  Neumann (periodic) boundary conditions
were applied in $x$- and $y$-directions in the surface plane for
simulations involving the gold surface, while zero boundary conditions
were applied in the perpendicular $z$-direction.  The simulation grid
was chosen such that at least $4$~\AA\ of space around the position of
each atom in the non-periodic directions was ensured.  The Au(111)
gold substrate was modeled by two layers of 54 atoms, each using the
experimental lattice constant of fcc gold of $a= 4.08$~\AA.  These
settings result in a rectangular unit cell of $26.0 \times
15.0$~\AA$^2$.  The Brillouin zone was sampled by $3 \times 3$
Monkhorst-Pack \cite{Monkhorst76prb} distributed $k$-points in the
periodic directions.

Potentials are scanned by fixing all gold atoms and a central atom of
PQP$^+$ (the nitrogen atom) and/or of ClO$_4^-$ (the chlorine atom) to
given positions while all other atoms were allowed to relax without
any symmetry constraints until all forces were below 0.05 eV/\AA.

\begin{figure}[htbp]
  \begin{center}
    \includegraphics[width=0.5\textwidth, clip]{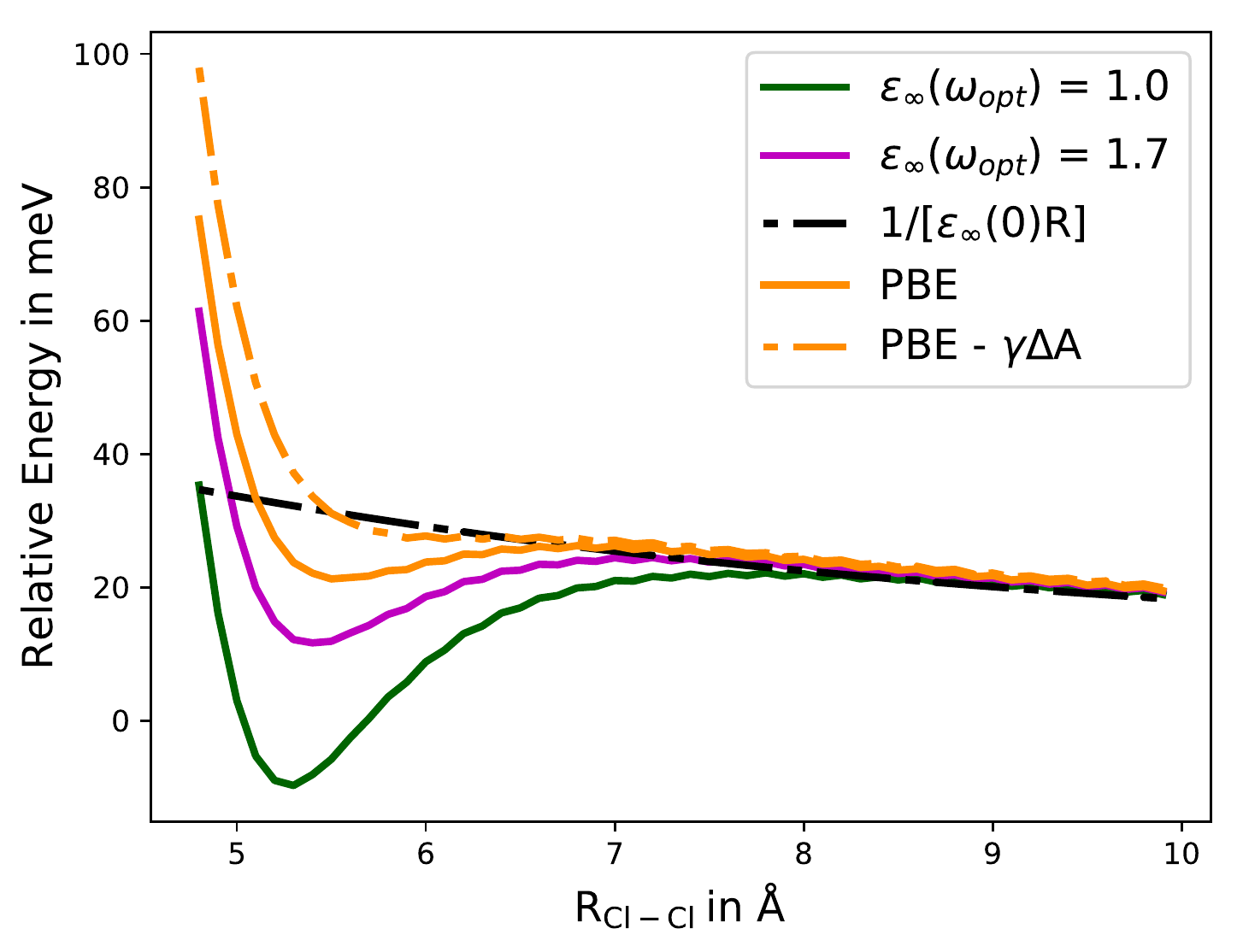}
    \caption{The relative energy of two ClO$_4^-$ anions as a function
      of the distance between their chlorine atoms, $R_{\rm Cl-Cl}$,
      where the separated anions define the energy reference;
      $\varepsilon(\omega_{\rm opt})=1$ with full van der Waals (vdW)
      corrections and $\varepsilon(\omega_{\rm opt})=1.7$ with scaled
      vdW corrections. The dash-dotted line shows the PBE energy
        where the energy contribution of effective surface tension
        $\gamma\Delta A$ is subtracted (see text).  }
    \label{fig:fit:2ClO4}
  \end{center}
\end{figure}

The interaction of two perchlorate anions in dependence of their
distance is shown in Fig.~\ref{fig:fit:2ClO4} for different
approximations for the total energy in Eq.~(\ref{eq:EDFT}).  As
expected, the potentials follow the screened electrostatic repulsion
$[\varepsilon_\infty(0)R_{\rm Cl-Cl}]^{-1}$ for large distances
$R_{\rm Cl-Cl}$, where $\varepsilon_\infty(0)=\varepsilon_r$ is the
static relative permittivity of water.  There is a slight attractive
part in the potential around $R_{\rm Cl-Cl}$ $\simeq 5.2$~\AA\ already
in the PBE potential which leads to a very shallow local minimum.
The main reason for this minimum is the decrease in
the effective surface $\Delta A$ when the solute cavities (the
solvent excluded regions) begin to overlap.  This decreases the
energetic cost to form the surface due to the effective surface
tension $\gamma=18.4$ dyn cm$^{-1}$ ($\gamma$ also contains
attractive contributions and is therefore much lower than the
experimental surface tension of water \cite{Held14jcp}). Subtracting
$\gamma\Delta A$ nearly removes all of the minimum as demonstrated
in Fig. ~\ref{fig:fit:2ClO4}.  Including the full dispersion
contribution [$\varepsilon_\infty(\omega_{\rm opt})=1$] this local
minimum substantially deepens and becomes the total minimum of the
potential.  An attractive contribution to the potential is not to be
expected for the interaction of two anions and needs further
discussion.  We suspect an overestimation of dispersion interactions
if these are treated as in vacuum and no screening through the
electrolyte is considered.

The aqueous environment influences the van der Waals (vdW)
interactions as these are of Coulombic origin \cite{Eisenschitz30zp}.
In order to derive an approximate expression for the screened vdW
interaction of two molecules $A$ and $B$ at distance $R$ inside the
electrolyte, we express the $C_6$ coefficient defining the vdW energy
$C_6/R^6$ by the Casimir-Polder integral
\cite{Tkatchenko09prl,Fiedler17jcpa}
\begin{equation}
  \label{eq:C6}
  C_6=\frac{3}{8\pi^2\varepsilon_0} \int_0^\infty \alpha^*_A(i\xi)\alpha^*_B(i\xi)
  \, \phi(i\xi) \, d\xi
\end{equation}
where $\alpha^*_{A,B}$ are the polarizabilities of the interacting
molecules and $\phi$ is determined by propagation of the electric
field through the embedding medium\cite{Fiedler17jcpa} with $\phi=1$
in vacuum. Both $\alpha^*_{A,B}$ and $\phi$ are modified relative to
vacuum in solution.  In the simplest model\cite{Fiedler17jcpa} we may
write $\phi(i\xi)=\varepsilon_\infty^{-2}(i\xi)$ with the frequency
dependent relative permittivity of the electrolyte
$\varepsilon_\infty$.  The effects of the electrolyte on the
polarizabilities $\alpha^*_{A,B}$ should, at least partly, already be
included in the TS09 description through the effective atomic
polarizabilities derived from the self-consistent electron density
calculated within the electrolyte.  What is left is the effect of the
permittivity entering through the function $\phi(i\xi)$ in
Eq.~(\ref{eq:C6}).  We assume that the main contribution of
$\phi(i\xi)$ is at the resonance frequencies of $\alpha^*_{AB}$, which
are in the optical region for usual molecules. We further assume that
$\varepsilon_\infty(\omega_{\rm opt})$ is approximately constant in
this frequency region, such that we may pull
$\phi=[\varepsilon_\infty(\omega_{\rm opt})]^{-2}$ out of the
integral.  This factor scales the $C_6$ coefficient and therefore the
vdW contribution. In other words, we apply the weight
$w_s=[\varepsilon_\infty(\omega_{\rm opt})]^{-2}$ in Eq.~(\ref{eq:C6})
with the experimental permittivity of water in the optical region of
$\epsilon_\infty(\omega_{\rm opt}) = 1.7$, see
Ref.~\citenum{Beneduci08jml}.  This approach reduces considerably the
depth of the suspiciously deep minimum as seen in
Fig.~\ref{fig:fit:2ClO4} such that only a shallow local minimum
remains similar to the PBE potential.  The reduction
obtained is quite strong in respect of the small contributions of
Axilrod-Teller-Muto interactions commonly
assumed\cite{Bukowski01jcp,McDaniel14jpcb}.
The quantitative connection between the screening of dispersive
interactions in polarizable media and the many-body effects neglected
in TS09\cite{Tkatchenko12prl} are not immediately clear and is
certainly worth further investigation.  In what follows we use the same
scaling for all the vdW contributions of the DFT potentials in this work.

\subsection{Force field model}
\label{subsec:system_classical}

In this subsection we describe how we cast our setup into
force fields where the atoms in the molecular
constituents are described as spherical particles, each of them
carrying a charge.  The mapping is guided by the energies obtained via
the {\it ab initio} simulations detailed above.

The Au(111) surface is modeled as a flat and perfectly conducting
surface involving mirror charges as detailed below.  However, we note,
that the position of the corresponding surface in the DFT calculations
does not coincide with the position of the atoms.  Before proceeding
he following comment is in order: in this mapping procedure the
distance of a point charge to a metallic surface is unambiguously
defined through the electrons leaking out of the potential defined by
the nuclei \cite{Perdew88prb,Serena86prb}. This feature can explicitly
be seen in jellium models \cite{Lang73prb}, but emerges also in
implicit calculations \cite{Finnis95jpc} where electrons spill out of
the surface of metal clusters \cite{Held13prb}. From the latter study
we estimate an effective spill out of the surface of 0.5 \AA{}, a
value that agrees qualitatively with estimates from the jellium
models, extrapolated to large structures \cite{Serena86prb}. This
value will be used in the following for our problem.

\subsubsection{The atomistic model}
\label{subsubsec:atomistic-model}

In our atomistic model the molecules are represented as rigid entities
composed of atomistic constituents. The molecules are immersed into a
microscopic electrolyte, which is treated as a continuous medium of
given permittivity. From below the system is confined by a conducting
Au(111)-surface (which is assumed to extend in the $x$- and
$y$-directions), an external field (with respect to the electrolyte)
can be applied in $z$-direction, i.e., perpendicular to the surface
(or wall).  Fig.~\ref{fig:experiment:atomistic} schematically depicts
all details of this atomistic model for the PQP$^+$ClO$_4^-$ system,
confined by the Au-surface.

In order to specify the different entities of the system and their
force fields we use the following notation:

\begin{itemize}
\item[(i)] Each of a total number of $\NumberOfMolecules$ molecules is
  uniquely labeled by capital Latin indices
  $\Molecule{I}$: for each of these units this index is assigned to
  its center-of-mass (COM) position vector, $\MoleculePosition{I}$, to
  a vector $\MoleculeOrientation{I}$, specifying its orientation
  within the lab-frame in terms of the angle-axis framework
  \cite{Chakrabarti2014, Chakrabarti2008} (see
  S.I.~Subsection~2.2 for more details),
  and to the set of coordinates, $\BlueprintMolecule{I}$, of the
  respective $\NumberOfConstituent{I}$ atomistic constituents of the
  molecule in its COM-frame (to which we also refer as its {\it
    blueprint}).  The set of COM-positions and orientation-vectors of
  all $\NumberOfMolecules$ molecules are denoted by $\BlueprintCOM$
  and $\BlueprintQuat$.  The set of all $\NumberOfAtomsFromMolecules$
  atom positions in the lab-frame is given by $\BlueprintLabframe$,
  and the position of each atom in the lab-frame is uniquely defined
  by a vector $\PositionVector{i}$, labeled with Latin indices ($i=1,
  \ldots, n$).

\item[(ii)] Between all atoms we consider long-range Coulombic
  interactions (index 'C'),

\begin{equation}
U^{\rm{(C)}}(r_{ij})= \frac{1}{4\pi \epsilon_0
  \epsilon_r}\frac{q_i q_j}{r_{ij}} ~~~~~ i \ne j
\label{eq:Coulomb}
\end{equation}
with the inter-atomic distance $r_{ij}=|\mathbf r_i - \mathbf r_j|$
and charges $q_i$ and $q_j$ of the units $i$ and $j$; the dielectric
constant $\epsilon_0$ and the relative permittivity $\epsilon_r$
specify the implicit electrolyte. Further, we introduce short-range
force fields (index 'sr') for which we have considered
two options: first, a Lennard-Jones potential (index 'LJ'), i.e.,

\begin{equation}
U^{\rm{(LJ)}}(r_{ij})= 4
\epsilon_{ij}\left[\left(\frac{\sigma_{ij}}{r_{ij}}\right)^{12} -
\left(\frac{\sigma_{ij}}{r_{ij}}\right)^{6}\right] ;
\label{eq:LJ}
\end{equation}
for the energy- and length-parameters, $\epsilon_{ij}$ and
$\sigma_{ij}$, we have opted for the standard Lorentz-Berthelot mixing
rules \cite{HansenMcDonald}, i.e.,
$\sigma_{ij}=\frac{1}{2}(\sigma_i+\sigma_j)$ and
$\epsilon_{ij}=\sqrt{\epsilon_i \epsilon_j}$, respectively.

Alternatively, we have also considered for the short-range
interactions the Mie potential \cite{Mie1903} (index 'Mie'), which can
be considered as a generalization of the LJ interaction; its
functional form is given by

\begin{equation}
U^{\mathrm ({\rm Mie})}(r_{ij}) = C_{ij} \epsilon_{ij} \left[
\left(\frac{\sigma_{ij}}{r_{ij}}\right)^{\MieRep{ij}}
- \left(\frac{\sigma_{ij}}{r_{ij}}\right)^{\MieAtr{ij}}
\right] ;
\label{eq:mie}
\end{equation}
and allows for a variation of the exponents of the repulsive and
attractive contributions to the potential, $\MieRep{ij}$ and
$\MieAtr{ij}$, respectively. $\epsilon_{ij}$ and $\sigma_{ij}$ are
again parameters for the energy- and the length-scales. The $C_{ij}$
are defined as functions of the exponents \cite{Mie1903}:

\begin{equation}
C_{ij}=\left(\frac{\MieRep{ij}}{\MieRep{ij}-\MieAtr{ij}} \right)
\left( \frac{\MieRep{ij}}{\MieAtr{ij}} \right)
^{\left(\frac{\MieAtr{ij}}{\MieRep{ij} - \MieAtr{ij}} \right)}    ;
\label{eq:mie:c_ij}
\end{equation}
for the exponents we apply arithmetic mixing laws, i.e.,
$\MieRep{ij}=\frac{1}{2}(\MieRep{i}+\MieRep{j})$ and
$\MieAtr{ij}=\frac{1}{2}(\MieAtr{i}+\MieAtr{j})$.

\item[(iii)] We assume the Au-surface to be perfectly conductive,
  consequently we need to explicitly consider mirror-charges in our
  model; when further assuming $z=0$ as the plane of reflection, the
  Coulombic interaction becomes

  \begin{equation}
U^{\prime\,\mathrm{(C)}} (r_{ij}) =
  U^{\mathrm{(C)}}(r_{ij}) +
  U^{\mathrm{(C)}}(r_{i j^\prime}) +
  U^{\mathrm{(C)}}(r_{i^\prime j}) +
  U^{\mathrm{(C)}}(r_{i^\prime j^\prime})
  \label{eq:coulomb:image}
\end{equation}
with the mirror charges $q_{i^\prime}=-q_i$ and their positions
$\mathbf r_{i^\prime}=(x_{i^\prime}, y_{i^\prime}, z_{i^\prime})=(x_i,
y_i, -z_i)$.

\item[(iv)] We describe the solid--liquid interface in terms of a
  slab-geometry with a lower confining wall, i.e., we assume
  periodicity in the $x$- and $y$-directions, but a finite extent,
  $c_z$, of the geometry in the $z$-direction which is chosen such
  that no restriction in the orientation of any molecule occurs, thus
  $c_z\approx1.2-2$~nm, given their size and the slab-width.  We
  define the (orthorhombic) lattice vectors, $\mathbf{a} = (a_x, 0,
  0)$, $\mathbf{b} = (b_x, b_y, 0)$, and $\mathbf{c} = (0, 0, c_z)$,
  which, without loss of generality, define the volume of the
  unit-cell, $V=a_x\,b_y\,c_z$, and which we collect within the matrix
  $\UnitCell=(\mathbf{a},\mathbf{b}, \mathbf{c})$.  Together with the
  molecular basis, given by $\BlueprintCOM$, $\BlueprintQuat$ and all
  $\NumberOfMolecules$ (rigid) molecular blueprints,
  $\BlueprintMolecule{I}$, we now define the supramolecular lattice

  \begin{equation}
    \Genome{}=\Lattice=
      \{\BlueprintCOM, \BlueprintQuat, \UnitCell\},
    \label{eq:lattice}
  \end{equation}
which gives rise to all atomic coordinates in the lab-frame,
$\BlueprintLabframe$, i.e., the molecular crystal structure of the
system (see S.I.~Subsection~2.2).

\item[(v)] The force field between the atomic entities
  and the Au-surface is described via a LJ-type wall potential
  \cite{Magda1985},

\begin{equation}
U^{({\mathrm{wall}})}(z_i) = 2\pi\epsilon_{{\rm w}i}\left[
\frac{2}{5}\left(\frac{\sigma_{{\rm w}i}}{z_i}\right)^{10}
- \left(\frac{\sigma_{{\rm w}i}}{z_i}\right)^4
- \frac{\sqrt 2 \sigma_{{\rm w}i}^3}
{3(z_i + (0.61/\sqrt 2) \sigma_{{\rm w}i})^3}\right] ;
\label{eq:LJ-wall:10-4-3}
\end{equation}
in the above relation, $z_i$ is the height of atom $i$ above the
surface, $\sigma_{{\rm w}i}$ and $\epsilon_{{\rm w}i}$ are the length-
and energy-parameters of the interactions of each atom $i$ with the
wall, respectively.

\item[(vi)] Finally, we express the electrostatic interfacial
  potential between the electrode and the Au-surface by an external,
  homogeneous electrostatic field, $E_z$ (i.e., oriented perpendicular
  to the surface): we account for this potential via
  $U^{(\rm{field})}(z_i) = z_i\,q_i\,E_z$ \cite{Jacob2015}.
\end{itemize}

Thus and eventually the total potential energy of our model is given
by the expression

\begin{equation}
\label{eq:model:atomistic}
U(\BlueprintLabframe, \UnitCell; E_z) =
\sum\limits_{i\neq j}^n{\vphantom{\sum}}^*
\left[U^{\prime\,\rm{(C)}}(r_{ij}) + U^{\rm{(sr)}}(r_{ij}) \right] +
\sum\limits_{i=1}^n\left[U^{\rm{(wall)}}(z_i) +
  U^{(\rm{field})}(z_i)\right],
\end{equation}
with 'sr' standing for 'LJ' or 'Mie'; we recall that
$\BlueprintLabframe$ is the set of all $\NumberOfAtoms$ atomic
positions $\PositionVector{i}$ in a lattice with slab-geometry
(defined by the unit-cell $\UnitCell$).  If not present (and not
explicitly addressed) the electric field will be dropped in the
argument list of Eq.~(\ref{eq:model:atomistic}), that is
$U(\BlueprintLabframe, \UnitCell; E_z=0) \equiv U(\BlueprintLabframe,
\UnitCell)$.  The notation '$\sum^*$' indicates that summation is only
carried out over atoms, labeled with Latin indices $i$
  and $j$, which belong to different molecules $I$ and $J$ (with
$I\neq J$); molecules being labeled with capital
  indices.  The energy given in Eq.~(\ref{eq:model:atomistic}) and
the corresponding force fields are efficiently evaluated using the
software-package LAMMPS \cite{Plimpton1995}.

To evaluate the long-range Coulomb term, $\sum\limits_{i\neq
  j}^n{\vphantom{\sum}}^*\,U^{\prime\,\rm{(C)}}(r_{ij})$ in the given
slab-geometry we use numerically reliable and efficient slab-corrected
3D Ewald-summation techniques \cite{Mazars2011, Smith1981, Yeh2009}.
The other terms in Eq.~(\ref{eq:model:atomistic}) are evaluated via
direct lattice summation techniques.

\FloatBarrier

\subsubsection{Parametrizing the classical model via ab-initio calculations}
\label{subsubsec:parametrizing}

In this work, the blueprint of each molecule $\BlueprintMolecule{I}$
is obtained from electronic structure calculations based on density
functional theory (DFT), using dispersion corrected {\it ab initio}
structure optimization \cite{Perdew96prl,Tkatchenko09prl}, as
described in Subsection~\ref{subsec:system_ab_initio}.  The partial
charges of the atoms, $q_i$, are parametrized via a Bader analysis
\cite{Bader1985} and are collected in Tables~2 and
3 in the S.I.~Subsection~2.3.  These
charges are directly transferred to the atomic entities.  We repeat
that throughout the electrolyte molecules have not been considered
explicitly: instead, we treat within the force field the
electrolyte as an effective, homogeneous medium, introducing the
electric permittivity of water $\epsilon_r$.

In order to fix the remaining model parameters that specify the
interactions in Eq.~(\ref{eq:model:atomistic}) we search for each
atomistic entity (labeled $i$) the set of atomistic model parameters
(specified below) which reproduces via Eq.~(\ref{eq:model:atomistic})
the {\it ab initio} energies as good as possible. On one side we
consider either the length- and the energy parameters of the
LJ-potential (denoted by ${\mathcal L} = \{ \sigma_i, \epsilon_i \}$)
or the length- and the energy parameters together with the exponents
of the Mie-potentials (denoted by ${\mathcal M} = \{ \sigma_i,
\epsilon_i, \MieRep{i}, \MieAtr{i} \}$), as well as the wall
parameters, $\mathcal{W} =\{\sigma_{{\rm w}i}, \epsilon_{{\rm
    w}i}\}$. To fix these parameters we proceed as follows:

\begin{itemize}
\item[(i)] We first perform {\it ab initio} structure optimization for
  different, characteristic molecular configurations, specified below.
  Here, molecules are either positioned next to each other (without
  considering the wall) or above the Au-surface: in the former case we
  fix the positions of two selected atoms belonging to different
  molecules, the atoms being separated by $r_{ij}$; in the latter case
  we keep the height, $z_k$, of one selected atom above the surface
  constant.  Relaxation of all other degrees of freedom leads in the
  {\it ab initio} simulations to spatially and orientationally
  optimized molecular structures; they are denoted by
  $\BlueprintDFT{n}{r_{ij}}$ and $\BlueprintDFT{n}{z_k}$,
  respectively, with corresponding energies
  $U_{\rm{DFT}}\left(\BlueprintDFT{n}{r_{ij}}, \UnitCell\right)$ and
  $U_{\rm{DFT}}^{(\rm wall)}\left(\BlueprintDFT{n}{z_k},
  \UnitCell\right)$; they are, themselves, functions of the
  inter-atomic distance, $r_{ij}$, and the atom-wall separation,
  $z_k$, of the selected atoms.

\item[(ii)] For every optimized {\it ab initio} structure,
  $\BlueprintDFT{n}{r_{ij}}$ and $\BlueprintDFT{n}{z_k}$, obtained in
  this manner we define a corresponding molecular configuration
  $\BlueprintFit{n}{r_{ij}}$ and $\BlueprintFit{n}{z_{k}}$, which is
  based on the above introduced rigid atomistic model
  $\BlueprintMolecule{I}$ (with the index $\Molecule{I}$ running now
  over all $\NumberOfMolecules$ molecules present in the respective
  DFT structure).  To this end we synchronize the COM-positions of
  each molecule $\Molecule{I}$ in the {\it ab initio} simulation with
  the corresponding COM-positions $\MoleculePosition{I}$ of its
  classical counterparts and align their orientation
  $\MoleculeOrientation{I}$ accordingly.

 \item[(iii)] Finally we evaluate the corresponding energies
   with the help of the force field via
   Eq.~(\ref{eq:model:atomistic}) at zero electric field,
   i.e. $U_{\mathcal{L/M}}\left(\BlueprintFit{n}{r_{ij}},
   \UnitCell\right)$ and $U^{(\rm
     wall)}_{\mathcal{L/M},\mathcal{W}}\left(\BlueprintFit{n}{z_{k}},
   \UnitCell\right)$.  We search for the best set of parameters
   $\mathcal{L}$ (or $\mathcal{M}$) and $\mathcal{W}$ via
   simultaneously minimizing

\begin{subequations}
\begin{align}
\label{eq:functional:molecules}
{\mathcal F}_{\mathcal L/M}&=\sum_{\{r_{ij}\}}
  \left|
  U_{\rm{DFT}}\left(\BlueprintDFT{n}{r_{ij}}, \UnitCell\right) -
  U_{\mathcal{L/M}}\left(\BlueprintFit{n}{r_{ij}}, \UnitCell\right)
  \right|^2 \\
\label{eq:functional:wall}
{\mathcal F}^{\rm(wall)}_{\mathcal L/M,\mathcal W}&=\sum_{\{z_k\}}
  \left|
  U_{\rm{DFT}}^{(\rm wall)}\left(\BlueprintDFT{n}{z_k}, \UnitCell\right) -
  U^{(\rm wall)}_{\mathcal{L/M},\mathcal{W}}\left(\BlueprintFit{n}{z_{k}}, \UnitCell\right)
  \right|^2.
\end{align}
\end{subequations}

Of course, in the model the same unit-cell, $\UnitCell$, and the same
number of particles, $\NumberOfAtoms$, as in the respective {\it ab
  initio} simulations have to be used. Note that in
Eq.~(\ref{eq:functional:molecules}) the wall-term included in
Eq.~(\ref{eq:model:atomistic}) is obsolete since the surface atoms are
not considered.
\end{itemize}

\begin{figure}[htbp]
\begin{center}
\includegraphics[width=0.85\textwidth]{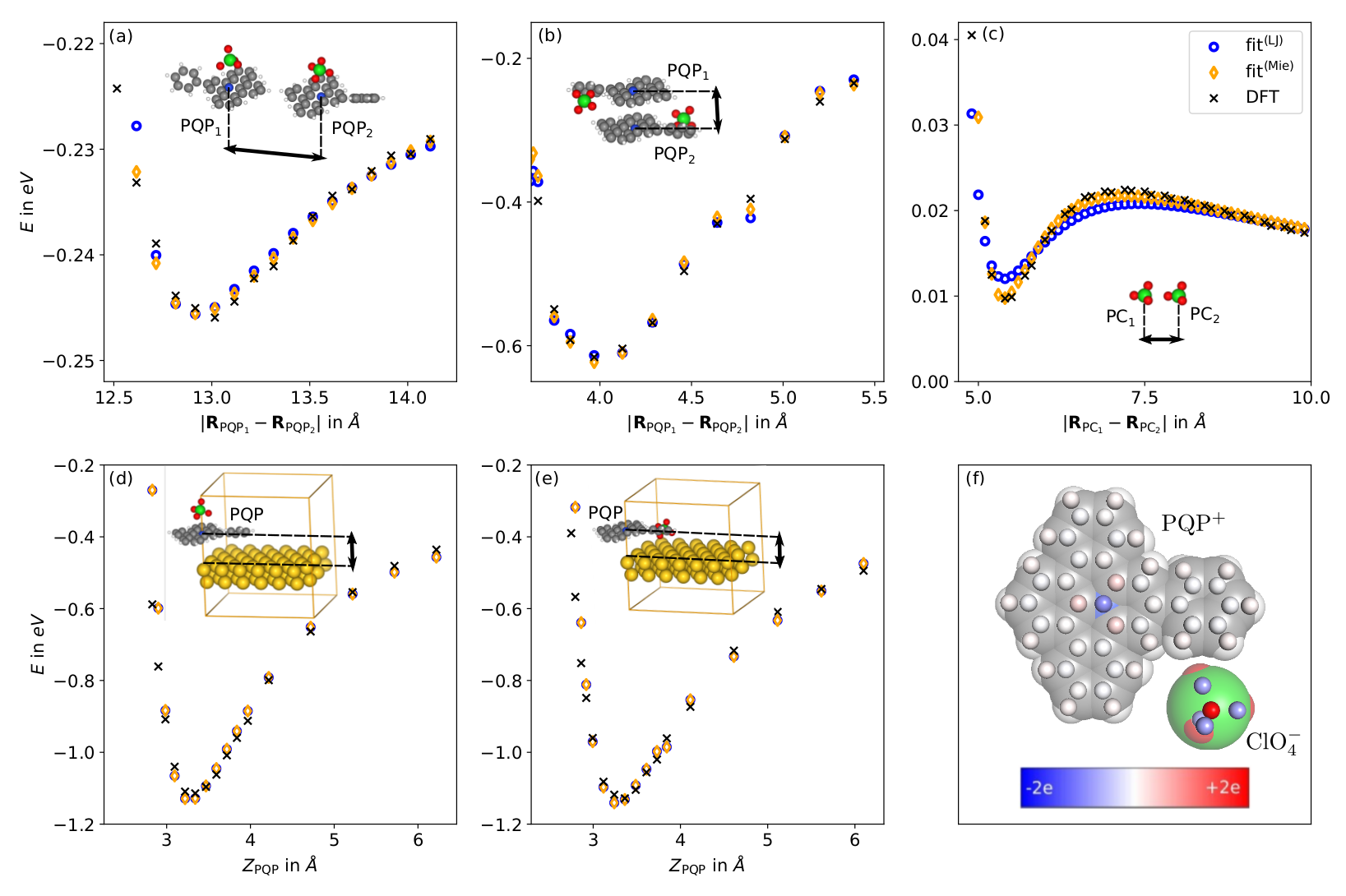}
\caption{(color online) Energies as obtained in {\it ab initio}
  simulations (black crosses) and fitted data, using the
  force field (involving a LJ interactions -- open blue
  circles -- or Mie interactions -- open orange diamonds), see
  Subsection~\ref{subsubsec:atomistic-model}; also shown are -- with
  labels (a) to (e) -- five schematic sketches of the five
  archetypical configurations of the molecules (along with their
  relative displacements, schematically indicated via the arrows as
  the distances vary along the abscissa); the related energy curves
  are used to fit the parameters of the force field, as
  outlined in the text; the labels correspond to the itemization (a)
  to (e) used in Subsection~\ref{subsubsec:parametrizing}.  Panel (f):
  PQP$^+$ and the ClO$_4^-$ molecules, drawn to scale and using the
  Mie force field for the short-range interactions:
  atomic entities are shown as transparent spheres with their
  diameters fixed by their respective optimized $\sigma_i$-values and
  their Bader charges (see color code).}
\label{fig:fit:atomistic}
\end{center}
\end{figure}

These fits are based on five particularly chosen, archetypical
configurations, to be discussed in the following. In the panels of
Fig.~\ref{fig:fit:atomistic} we display schematic sketches of these
configurations of the PQP$^+$ and ClO$_4^-$ molecules; these panels
show the corresponding energy curves obtained from the force field,
with parameters based on a fitting procedure to the {\it ab initio}
energy profiles.

\begin{itemize}
\item[(a)] {\it Tail-to-tail} configuration (see inset of in panel (a)
  in Fig.~\ref{fig:fit:atomistic}): We have considered a series of
  {\it ab initio} structure optimizations at constant, but
  successively increasing nitrogen-nitrogen distances, $r_{\rm NN}$,
  in the $x$-direction (while keeping $y_{\rm NN}$ and $z_{\rm NN}$
  constant) of an anti-parallel oriented pair of PQP$^+$ molecules;
  both cations are vertically decorated with a ClO$_4^-$ molecule.
  The aromatic parts of the PQP$^+$ molecules lie flat in the $x$- and
  $y$-directions such that their tails face each other.
\item[(b)]{\it Face-to-face} configuration (see inset in panel (b) in
  Fig.~\ref{fig:fit:atomistic}): In this case we consider
  anti-parallel oriented, but vertically stacked PQP$^+$ molecules
  (both being horizontally decorated by ClO$_4^-$ molecules) under the
  variation of the nitrogen-nitrogen distance, $r_ {\rm NN}$, in
  $z$-direction (while now keeping $x_{\rm NN}$ and $y_{\rm NN}$
  constant).  Again, the aromatic parts of the PQP$^+$ molecules lie
  flat in $x$- and $y$-directions; however, and in contrast to case
  (a) these units face each other.
\item[(c)]{\it ClO$_4^-$--ClO$_4^-$} configuration (see inset in panel
  (c) in Fig.~\ref{fig:fit:atomistic}): Here two ClO$_4^-$ molecules
  are considered, varying the chlorine-chlorine $x$-distance, $r_{\rm
    ClCl}$, while keeping $y_{\rm ClCl}$ and $z_{\rm ClCl}$ constant.
\item[(d)]{\it Face-to-wall topped} configuration (see inset in panel
  (d) in Fig.~\ref{fig:fit:atomistic}): In this case a single PQP$^+$
  molecule, lying flat and parallel to the $(x, y)$-plane, is located
  above two layers of Au and is vertically decorated by a ClO$_4^-$
  molecule. The cell geometry is assumed to be periodic in the $x$-
  and $y$-directions and finite along the $z$-axis; in an effort to
  scan along the $z$-direction, we have performed a series of {\it ab
    initio} based structure optimizations for selected fixed values of
  $z_{\rm N}$, i.e., the $z$-position of the nitrogen in PQP$^+$ above
  the Au-surface. The LJ 10-4-3 potential \cite{Magda1985} has been
  used between the Au(111) surface and the molecules.
\item[(e)]{\it Face-to-wall beside} configuration (see inset in panel
  (e) in Fig.~\ref{fig:fit:atomistic}): In contrast to case (d), the
  PQP$^+$ cation is now horizontally decorated by the ClO$_4^-$ anion
  such that both molecules are adsorbed on the Au-surface. Again, the
  LJ 10-4-3 potential \cite{Magda1985} has been used between the
  Au(111) surface and the molecules.
  \end{itemize}

In practice we first optimize ${\cal F}_{\cal L/M}$, given in
Eq.~(\ref{eq:functional:molecules}), involving thereby {\it all}
inter-atomic force field parameters; their values are
listed in Table~\ref{tab:fit:atomistic:vdw} for the LJ and the Mie
models. These parameters are then kept fixed and are used in the
subsequent calculations to optimize the wall force field
parameters via optimizing ${\cal F}^{\rm(wall)}_{{\cal L/M}, {\cal
    W}}$, specified in Eq.~(\ref{eq:functional:wall}); the emerging
parameters are listed in Table~\ref{tab:fit:atomistic:wall}.  In panel
(f) of Fig.~\ref{fig:fit:atomistic} we present a visualization of the
molecules PQP$^+$ and ClO$_4^-$, using these optimized parameters and
providing information about the charge of the atomic entities via the
color code.

\section{Identifications of self-assembly scenarios}
\label{sec:self_assembly}

With the classical force field for the PQP$^+$ and
ClO$_4^-$ molecules introduced in
Subsection~\ref{subsubsec:atomistic-model} at hand we are now ready to
identify the ordered ground-state configurations of these molecules as
they self-assemble on the Au-surface -- immersed into an electrolyte
and exposed to an electric field.  While we leave a more comprehensive
and systematic investigation of these self-assembly scenarios to a
future publication \cite{Hartl:prep-a}, we focus in this contribution
on the technical details of our approach and on a few selected sets of
external parameters (i.e, the electric field strength and the particle
density).

Our overall objective is to find for our system the global minimum of
the total free energy, $F$, at $T=0$~K as a function of the positions
and orientations of all molecules per unit-cell for a given value of
cell volume and $E_z$; at $T = 0$ this task reduces to the
minimization of the internal energy $U$. The minimum has to be found
in a huge-dimensional parameter space, spanning the positions and
orientations of the molecules and by the parameters specifying the
unit cell. To be more specific, the dimensionality is set
by the number of parameters to be optimized, which read 64, 76, and
88 for five, six, and seven molecules per unit cell, respectively.
It is a particular strength of our optimization algorithm (as
detailed in the following) to identify in an efficient and reliable
manner minima in such high dimensional search spaces.

For this purpose we use a memetic search algorithm which combines
evolutionary search strategies (EA)
\cite{Gottwald2005,Fornleitner2008a,Fornleitner2009,Doppelbauer2010,Fornleitner2010,Doppelbauer2012,Antlanger2016}
and local, steepest gradient descent procedures (LG)
\cite{Jones2001,Kraft1988}: First a total number of $\NumberOfGenomes$
different lattice-configurations, $\Genome{}=\Lattice$ as defined in
Eq.~(\ref{eq:lattice}), is generated. We note that
among those configurations we have also intentionally included as
``educated guesses'' molecular configurations, inspired by the
experimental self-assembly scenarios identified in
Ref.~\citenum{Cui14acie}; however it should be emphasized that this
information is only available for the PQP$^+$ ions, as the
experiment does not provides any information about the positions of
the perchlorate ions. This {\it population}, $\Population$, is
exposed to concepts of natural (or, rather, artificial) selection.  At
every iteration step of the EA a new configuration, i.e. an {\it
  offspring}, is created from existing configurations of the most
recent population, via crossover and mutation operation. This new
configuration is then subjected to an LG optimization, an operation
which represents by far the most time consuming task in our algorithm
and is performed in parallel using the ``{\it mpi4py}`` framework
\cite{Dalcin2005,Dalcin2008,Dalcin2011}.  For an optimal load-balance
we additionally spawn a master-thread on one of the mpi-processes to
asynchronously distribute optimization tasks of offspring
configurations among all idle mpi-processes.  The relaxed
configurations are gathered by this master-thread, which then decides
-- via a criterion primarily based on the respective internal energy
of the configurations -- whether the new relaxed particle arrangements
are accepted or rejected.

Since the experimental observations \cite{Cui14acie,Cui17small}
provide evidence of a structural organization of the molecules into
supramolecular lattices, the center-of-mass coordinates of the
molecules, $\BlueprintCOM$, and their orientations, $\BlueprintQuat$,
as well as the parameters defining the unit-cell, $\UnitCell$, (see
Subsection~\ref{subsubsec:atomistic-model} and
Fig.~\ref{fig:experiment:atomistic} for details), are the variables
which have to be optimized for the search of ground-state
configurations: We minimize $U(\BlueprintLabframe, \UnitCell; E_z)$,
defined in Eq.~(\ref{eq:model:atomistic}), with respect to
$\BlueprintCOM$, $\BlueprintQuat$, and $\UnitCell$, keeping the number
of molecules $\NumberOfMolecules$, the unit-cell volume $V$ (with
fixed slab width $c_z$), and the electrostatic field strength $E_z$
constant.

In more detail, we proceed as follows:

\begin{itemize}
\item[(i)] It is common in evolutionary algorithms to define a genome
  representation of the entity which is subject to optimization
  \cite{Mascato1989,GottwaldDieter2005}.  In our case we represent a
  supramolecular lattice configuration phenotypically (rather than
  genotypically \cite{GottwaldDieter2005}) by the set,
  $\Genome{}=\{\BlueprintCOM, \BlueprintQuat, \UnitCell\}$ as defined
  by Eq.~(\ref{eq:lattice}), i.e. the set of all COM coordinates and
  orientations of all molecules as well as the lattice vectors.

\item[(ii)] In the first step, two configurations (labeled
  henceforward by Latin indices), $\Genome{i}$ and $\Genome{j}$, are
  chosen at random or via the ''roulette wheel'' method (see item (v)
  below) from the evolutionary population
  \cite{GuentherDoppelbauer2012,Gottwald2005,Fornleitner2008,Fornleitner2008a,MoritzAntlanger2015};
  this strategy favors parents of high quality hence making them more
  likely to be used for reproduction than ''weaker'' configurations,
  i.e. configurations with higher energy from the evolutionary
  population.  Then these two configurations are combined via a
  crossover operation (i.e., a cut-and-splice process -- for more
  details see below) creating thereby an offspring configuration,
  $\Genome{i\oplus j}$, with the subscript '$i \oplus j$' emphasizing
  the executed crossover operation between $\Genome{i}$ and
  $\Genome{j}$.  supramolecular lattice.  The purpose of this
  operation is to save high quality blocks of the genetic material
  (e.g. the relative positions and orientations of molecules within
  the unit-cell) in order to efficiently sample the parameter space
  \cite{Mascato1989,GottwaldDieter2005,GuentherDoppelbauer2012,Gottwald2005,Fornleitner2008,Fornleitner2008a,MoritzAntlanger2015}.

\item[(iii)] In the second step, the newly generated offspring
  configuration, $\Genome{i\oplus j}$, is then exposed to random
  mutation moves: these are either translations or rotations of single
  molecules, swaps of center-of-mass positions or orientations of
  pairs of molecules or deformations of the unit-cell, each of them
  with a certain probability and within preset numerical boundaries.
  This step of the algorithm has the purpose of exploring disconnected
  areas in parameter space, a feature which is indispensable in global
  minimization techniques.

\item[(iv)] After these two steps, and assuming that the offspring
  configuration, $\Genome{i\oplus j}$, does not represent a local
  minimum with respect to the potential energy, a local energy
  minimization is performed. Here we mainly rely on the ``{\it
    scipy}'' implementation of the {\it SLSQP} gradient-descent
  algorithm \cite{Jones2001,Kraft1988} (allowing us to define
  numerical boundaries and constraints on the parameters during the
  optimization), which minimizes the forces and torques between the
  molecules as well as the stress of the unit-cell.  These tools are
  very helpful to keep the unit-cell volume fixed and to prevent
  re-orientations of the molecules where some of their atomic
  constituents would be transferred into positions outside the slab
  geometry, ensuring thus that $z_i>0$ for all atoms.  Subsequently we
  perform several ``basin-dropping`` (BD) steps, where we further try
  to lower the energy of the configuration by applying several small
  random ``moves`` in the parameter space of the LG-optimized
  offspring; from the emerging configurations only the ones with low
  energies are accepted.  This specific operation turned out to
  considerably improve the convergence rate of the local optimization,
  in particular if multiple and alternating sequences of LG and BD
  runs are applied.

\item[(v)] After the local search procedure the optimized offspring
  configuration, $\Genome{i\oplus j}$, becomes a new candidate to
  enter the evolutionary population, $\Population$.  The objective of
  the EA is to retain the best configurations (i.e., the energetically
  most favorable ones) within the population and to include only
  candidates with energy values better or comparable to those of the
  current population.  In an effort to quantify the quality of the
  candidates, their so-called fitness is evaluated
  \cite{GottwaldDieter2005, Gottwald2005, GuentherDoppelbauer2012,
    MoritzAntlanger2015, Fornleitner2008, Fornleitner2008a}, for which
  we have used in this contribution the function:

\begin{equation}
    F(U) = \exp\left(-s\dfrac{U-U_{\rm min}}{U_{\rm max} - U_{\rm min}}\right) ;
    \label{eq:fitness}
\end{equation}
$F(U)$ is a monotonic function of the energy $U$ of the candidates,
whose value ranges within the interval $0 \leq F(U) \leq F(U_{\rm
  min}) = 1$; $U_{\rm min}$ and $U_{\rm max}$ are the minimal and
maximal energies appearing in the population.  The selection parameter
$s$ quantifies the reproduction-rate for configurations within the
population in the sense that large values of $s$ tend to exclude
configurations with low fitness from reproduction; following
\cite{MoritzAntlanger2015} we commonly use $s=3$.  The aforementioned
``roulette wheel'' method for choosing suitable parent configurations
also relies on the fitness function (and hence the selection
parameter): Assuming that the configurations within the population
$\Population$ are sorted by their respective fitness values in
descending order, $F(U_i)>F(U_{i+1})$, the probability, $f(U_i)$, of a
configuration, $\Genome{i}$, to be selected for reproduction is given
in terms of the relative fitness \cite{Gottwald2005,
  GuentherDoppelbauer2012, MoritzAntlanger2015, Fornleitner2008}:

\begin{equation}
    f(U_i)=
      \sum\limits_{j=i}^{\NumberOfGenomes}F(U_j)
      \cdot\left[
      \sum\limits_{k=1}^{\NumberOfGenomes}\sum\limits_{j=k}^{\NumberOfGenomes}F(U_j)
      \right]^{-1},
    \label{eq:fitness:relative}
\end{equation}
$\NumberOfGenomes$ being the total number of configurations within
the population.  With a certain probability (commonly in $20\%$ of all
crossover moves) we allow reproduction between randomly chosen
configurations.

\item[(vi)] Once a new configuration is accepted to enter the
  population another configuration has to be eliminated.  The
  probability $p(U_i)$ for a configuration, $\Genome{i}$, to be
  replaced is given by

  \begin{equation}
    p(U_i)=
    \exp{[-s F(U_i)]}
    \left[
    \sum\limits_{j=1}^{\NumberOfGenomes}\exp{[-s F(U_j)]}
    \right]^{-1},
    \label{eq:extinction}
  \end{equation}
  a value which is again related to the fitness of the
  configuration, $F(U_i)$, and the selection parameter $s$.  Thus,
  configurations with low fitness are more likely to be eliminated.
  In any case, a few of the best configurations within the population
  are retained in an effort to keep the so far best solutions as
  appropriate parent candidates for the above-mentioned crossover
  procedures (a strategy referred to in the literature as {\it
    elitism} \cite{GuentherDoppelbauer2012}).

 It should be emphasized that this strategy does not follow biological
 selection mechanisms \cite{Darwin1859}, where populations are
 replaced entirely once that new generations have been formed;
 however, our strategy ensures to protect the best genetic material
 from extinction during the entire search procedure
 \cite{GuentherDoppelbauer2012,MoritzAntlanger2015}.

\item[(vii)] In an effort to maintain diversity within the population
  an additional operation (in literature referred to as {\it
    nichening} \cite{Hartke1993,Hartke1999}), is applied: locally
  optimized offspring configurations will be discarded if the values
  of their energy is too close to the energy of any configuration
  currently in the population; avoiding thereby that the population is
  overrun by structurally identical configurations. At the same time
  the maintenance of genetic diversity is guaranteed.

  However, this procedure alone cannot cope with 'degenerate'
  configurations, i.e., if structurally distinct configurations have
  essentially the same energy values (within the specified nichening
  tolerance).  In our approach we allow configurations to enter the
  population only if their structures differ significantly from those
  of the competing, degenerate configurations.  In order to quantify
  the structural difference between configurations we associate a
  feature vector, $\mathbf{f}_i$ (i.e. a set of order parameters),
  which collects a set of order parameters pertaining to configuration
  $\Genome{i}$ (see S.I.~Subsection~3.2 for
  details).  The degree of similarity between two configurations,
  $\Genome{i}$ and $\Genome{j}$, is then evaluated by taking the
  Euclidean distance between the corresponding feature vectors, i.e.,
  $\Delta_{ij}=|\mathbf{f}_i - \mathbf{f}_j|$; similar configurations
  will have a small distance, while unlike configurations will have a
  large distance.  If $\Delta_{ij}$ is above a certain threshold
  value, the offspring configuration, $\Genome{i \oplus j}$, will not
  be discarded by the energy-nichening operation.
  \end{itemize}

  In order to offer the reader an insight into the
  computational complexity of our project we outline via a few
  characteristic numbers the computational limitations: the bottleneck
  of the identifications of self-assembly scenarios are (i) the huge
  number of calls of energy-evaluations in the optimization step, as
  underlined via some example: per generation we have at least 10$^4$
  calls of the energy kernel; for each state point we need at least
  10$^4$ generations, which leads to an absolute minimum of 10$^8$
  calls of the energy kernel for \textit{one (!)} set of system
  parameters; (ii) the optimization of the energy in a high
  dimensional search space (as specified above), ranging from $\sim60$
  to $90$, depending on the number of molecules.

Summarizing, the complexity of the problem at hand forces us to use
all the above listed advanced optimization tools, including a basin
hopping memetic approach combining the heuristic nature of
evolutionary strategies with deterministic local gradient descent
algorithms \cite{Gottwald2005}.  The gradient descent method
deterministically evaluates every local minimum of the basin with high
accuracy (which is additionally sped up by the ''basin dropping''
procedure) while the evolutionary search gradually adapts its
population to the energetically most favorable solution, exploring the
search space for the global optimum.

  To round up this section it should be noted that a
  variety of techniques have been used in literature for related
  optimization problems; among those are: Monte Carlo or molecular
  dynamics-based techniques such as simulated
  annealing\cite{Pannetier1990,Schoen2010}, basin-hopping
  \cite{Wales1997,Panosetti2015,Krautgasser2016}, minima
  hopping\cite{Goedecker2004,Goedecker2010}, and eventually
  evolutionary approaches such as genetic algorithms
  \cite{Gottwald2005,Fornleitner2008a,Fornleitner2009,Doppelbauer2010,Fornleitner2010,Doppelbauer2012,Antlanger2016,Abraham2006,Deaven1995,Hartke1999,Hartke1993,Lyakhov2010,Supady2015}.
  The decision on the method of choice relies on the specific problem:
  for instance, as Hofmann {\it et al.} used the SAMPLE technique (see
  Refs.~\citenum{Hofmann2017,Hofmann2018,Hofmann2019}), relying on a
  discretization of the search space into limited, archetypical,
  intermolecular motives and elaborate data fitting of emerging force
  fields to describe intermolecular interactions.  To the best of our
  knowledge, this approach has neither been applied to molecular
  motives beyond monolayer configurations or to charged molecules, so
  far, nor has it been used in combination with an external control
  parameter, such as an electric field or systems composed of multiple
  components.  In general, the fact that the number of archetypical
  inter-molecular motives grows rapidly with the increasing size of
  the molecules bears the risk of hitting very soon the limits of
  computational feasibility.  However, suitable adaptations of this
  strategy and/or a combination with evolutionary search strategies or
  with reinforcement learning -- which has, for instance, very
  successfully been applied to protein folding problems
  \cite{AlphaFold} in a similar way as AlphaGo\cite{AlphaGo} was able
  to master the infamous board game -- might represent a viable route
  to circumvent the aforementioned limitations; thus future
  investigations of such intricate problems as the complex monolayer
  to bilayer transition, addressed in this contribution, might come
  within reach.

\section{Results}
\label{sec:results}

\subsection{General remarks and system parameters}
\label{subsec:results_general}

In the following we present selected results for self-assembly
scenarios of PQP$^+$ and ClO$_4^-$ molecules on an
Au(111)-electrolyte--interface under the influence of an external
electrostatic field, as obtained via the algorithm presented in the
preceding sections. Our choice of parameters is guided by the
experimentally observed molecular configurations \cite{Cui14acie}. We
demonstrate that our proposed strategy is indeed able to reproduce on
a semi-quantitative level the experimentally observed self-assembly
scenarios \cite{Cui14acie}.  As a consequence of the still sizable
costs of the numerical calculations we leave more detailed
investigations (where we systematically vary the system parameters)
and a quantitative comparison of our results with the related
experimental findings \cite{Cui14acie} to a future contribution
\cite{Hartl:prep-a}.

To be more specific we have used the following values for the
(external) system parameters:

\begin{itemize}
\item an indication for the number of molecules per unit cell is
  provided by the experiment \cite{Cui14acie}: we have considered unit
  cells containing ten, twelve, and 14 pairs of PQP$^+$ and ClO$_4^-$
  molecules. These numbers in molecules include, of course, also the
  related mirror molecules and correspond to 630, 636, and 742 atomic
  entities per unit cell, respectively (which interact via short-range
  and long-range potentials, which are subject to particle wall
  interaction and which are sensitive to an external electrostatic
  field);
\item also the actual values of the surface area $A$ is motivated by
  estimates taken from experiment \cite{Cui14acie}: we have varied $A$
  within the range of $6.5$~nm$^{2}$ to $12.25$~nm$^{2}$, assuming a
  step size of typically $0.5$~nm$^{2}$; systems will be characterized
  by the surface density of the PQP$^+$ molecules, defined as
  $\sigma_{\rm PQP} = N_{\rm PQP}/A$, $N_{\rm PQP}$ being the number
  of ${\rm PQP}$ molecules per unit cell;
\item the range of the experimentally realized values for the
  electrostatic field strength $E_z$ is, however, difficult to
  estimate since the major drop in voltage occurs near the negatively
  charged Au-surface and the nearby layers of cations
  \cite{Jacob2015}, which is not directly accessible in experiment.
  Therefore we have covered -- at least in this first contribution --
  several orders of magnitude in the value for $E_z$ within a range
  that extends (on a logarithmic grid) from $E_z=-1$~V/nm to
  $E_z=-10^{-3}$~V/nm; in addition, we have also performed
  calculations at zero electrostatic field.
\end{itemize}

It should be mentioned that we have used in all these calculations the
Mie potential within the classical model, since the related LJ model
is not able to fit the {\it ab initio} data with a comparable and
sufficient accuracy (see also discussion in
Subsection~\ref{subsubsec:atomistic-model}).

We have covered in total approximately $176$ combinations of these
parameters (that is the unit-cell volume $V$ with a constant slab
width $c_z$, the number of molecules $\NumberOfMolecules$, and the
electrostatic field strength $E_z$); for each of these we performed
independent evolutionary searches with a population size of typically
$\Population=40$ configurations.  Some details about the numerical
costs of our calculations can be found in
S.I.~Subsection~4.1

\subsection{Discussion of the results}
\label{subsec:results_discussion}

\subsubsection{Lateral particle arrangements}
\label{subsubsec:results_lateral}

\begin{figure}[htbp]
\begin{center}
\includegraphics[width=0.495\textwidth, clip=True]{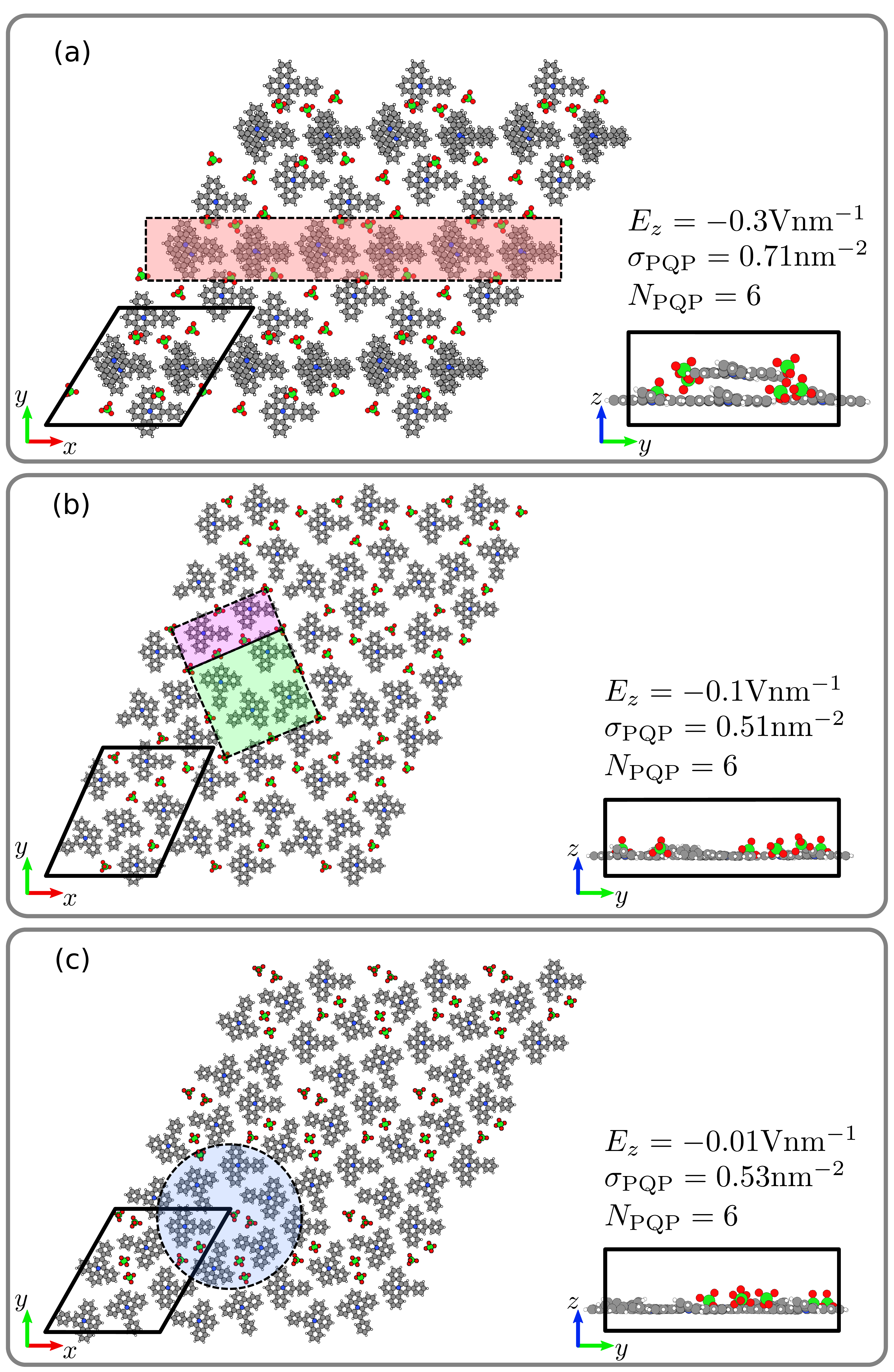}
\includegraphics[width=0.495\textwidth, clip=True]{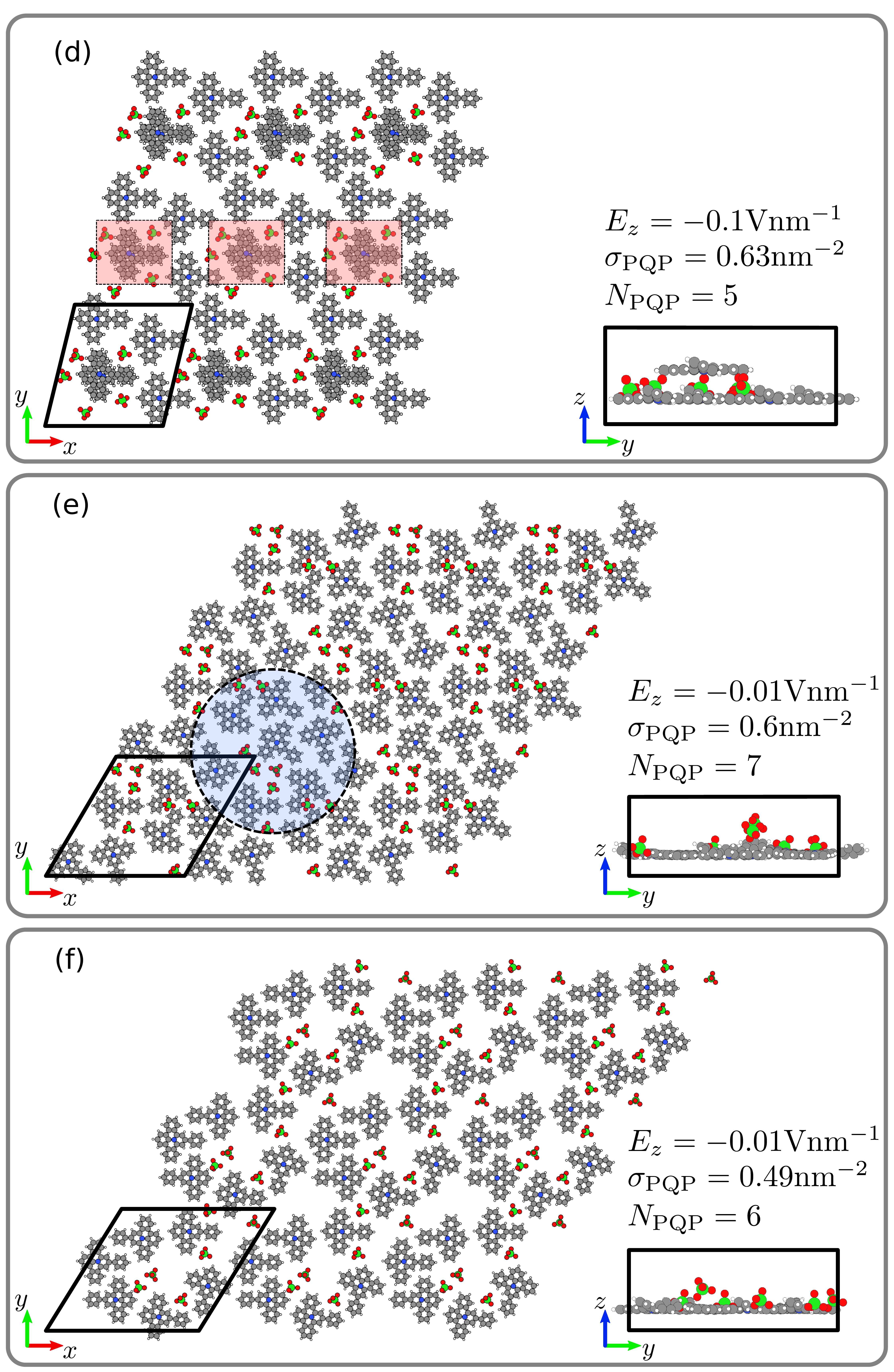}
\caption{(color online) Results for the ground state configurations of
  PQP$^+$ and ClO$_4^-$ molecules, adsorbed on a Au(111) surface under
  the influence of an external electrostatic field $E_z$, as they are
  obtained via the numerical procedure, as specified in
  Section~\ref{sec:self_assembly}; calculations are based on the
  classical model for the molecules, involving the Mie potential (for
  details see Subsection~\ref{subsubsec:atomistic-model}). In the main
  panels configurations are shown in a periodically extended view as
  projections onto the $(x, y)$-plane and in the respective insets as
  projections onto the $(y, z)$-plane; in the main panels the
  respective unit cells are highlighted by thick black lines. Results
  are shown for different values of the number of ${\rm PQP}$
  molecules, the surface density $\sigma_{\rm PQP}$ and the
  electrostatic field $E_z$: see labels in the different panels and
  Tab.~\ref{tab:results} for details.  The red (gray)
    shaded areas, framed with dashed lines, in panel (a) and (d)
    emphasize PQP$^+$ molecules which sit on top of other cations,
    forming a bilayer structure.  The dashed, shaded, magenta (gray)
    rectangular and green (gray) square areas in panel (b) represent
    tilings formed by perchlorate molecules within the dense PQP$^+$
    monolayer configuration.  The dashed, shaded, blue (gray) circles
    in panel (c) and (e) emphasize, quantitatively, the porous and
    auto-host--guest tiles identified in experiment (see Fig.~4A and
    Fig.~4C in Ref.~\citenum{Cui14acie}, respectively).}
\label{fig:ea:results_experimental_trend}
\end{center}
\end{figure}

In this subsection we discuss the lateral self-assembly
scenarios of the PQP$^+$ and of the ClO$_4^-$ molecules.  Selected
results for our numerical investigations are presented in
Fig.~\ref{fig:ea:results_experimental_trend} and in -- on a more
quantitative level -- in Table~\ref{tab:results}. The actual values
have been chosen in an effort to reproduce -- at least on a
qualitative level -- the results obtained in the experimental
investigations. Indeed, the sequence of the obtained ordered ground
state configurations (shown in panels (a) to (c)) clearly indicates
the transition from a stratified bilayer configuration (identified at
a rather strong electrostatic field strength of $E_z=-0.3$~V/nm), over
a self-hosts--guest mono-layer structure (obtained by reducing the
field down to $E_z=-0.1$~V/nm), and eventually to an open-porous
configuration (identified at $E_z=-0.01$); similar observations have
been reported in the related experimental study \cite{Cui14acie}.

From the results of our investigations (which are shown only
selectively) we learn that an electrostatic field strength of
$E_z=-0.3$~V/nm always leads to bilayer configurations, similar to the
one shown in panel (a) of
Fig.~\ref{fig:ea:results_experimental_trend}.  This stratified bilayer
configuration represents the energetically most favorable one as we
vary at fixed $E_z$ the volume of the unit cell and the number of
molecules within the respective ranges, specified in the preceding
section; the numerical data of the related internal energy are
compiled in Table~\ref{tab:results}.

As we proceed to $E_z=-0.1$~V/nm we observe self-assembly scenarios as
the ones depicted in panels (b) and (d) of
Fig.~\ref{fig:ea:results_experimental_trend}, which correspond to
self-hosts--guest configurations observed in experiment
\cite{Cui14acie}; for the data presented in these panels two different
values for $N_{\rm PQP}$ (and hence for $\sigma_{\rm PQP}$) have been
considered: the mono-layer configuration shown in panel (b) has a
slightly better value for the internal energy (per molecule) than the
rhombohedral bilayer configuration shown in panel (d); however, as
can be seen from Table~\ref{tab:results} the energy differences are
very tiny: differences of the order of 10$^{-4}$eV correspond to
values where we hit the numerical accuracy of the {\it ab initio}
based energy values).

Eventually, we arrive at the so-called open-porous structures,
observed in experiment \cite{Cui14acie}: the ground state
configurations depicted in panels (c), (e), and (f) of
Fig.~\ref{fig:ea:results_experimental_trend} are evaluated at the same
electrostatic field strength of $E_z=-0.01$~V/nm, assuming different
values for $N_{\rm PQP}$ and $\sigma_{\rm PQP}$; the open-porous
pattern emerging in panels (c) is the most favorable one in terms of
energy per molecule (see Table~\ref{tab:results} for the numerical
details).  There are, however, several serious competing structures
with minute energy differences at this value of the electric field
strength: another open-porous structure, depicted in panel (f), with
an energy penalty of less than $8.1$~meV per PQP$^+$ molecule compared
to case (c) and also a considerably denser configuration, depicted in
panel (e), with an internal energy value worse by only $11.3$~meV
compared to case (c) and by $3.2$~meV compared to case (f).

From the numerical point of view the following comments
are in order: for a fixed state point, the energy differences of
competing structures attain values which hit the limits of the
accuracy of the {\it ab initio} based simulations, which can be
estimated to be of the order of 0.1 eV to 0.01 eV
per molecule for dispersive interactions
\cite{Tkatchenko09prl,Hanke11jcc,Berland15rpp,Stohr19csr}.
These values set the limits of our numerical accuracy.
For completeness we note that for the results for the energies
obtained via the classical force field (which are based on LAMMPS
calculations) we estimate that our results are numerically reliable
down to $\sim 10^{-6}$ per atom; within the range
of such minute energy differences no competing structures have been
found in our investigations.
In general we observe that the energy differences for the
energetically optimal ground state configurations become smaller as
the electrostatic field tends towards zero. Even though the
optimization algorithm (as outlined in Section~\ref{sec:self_assembly})
has turned out to be very efficient and reliable, we observe (in
particular for smaller values of the external field)
that new configurations are included in the population of the best
individuals even after a large number of optimization steps.

\subsubsection{Vertical particle arrangements}
\label{subsubsec:results_vertical}

\begin{figure}[htbp]
\begin{center}
\includegraphics[width=\textwidth,
  clip=True]{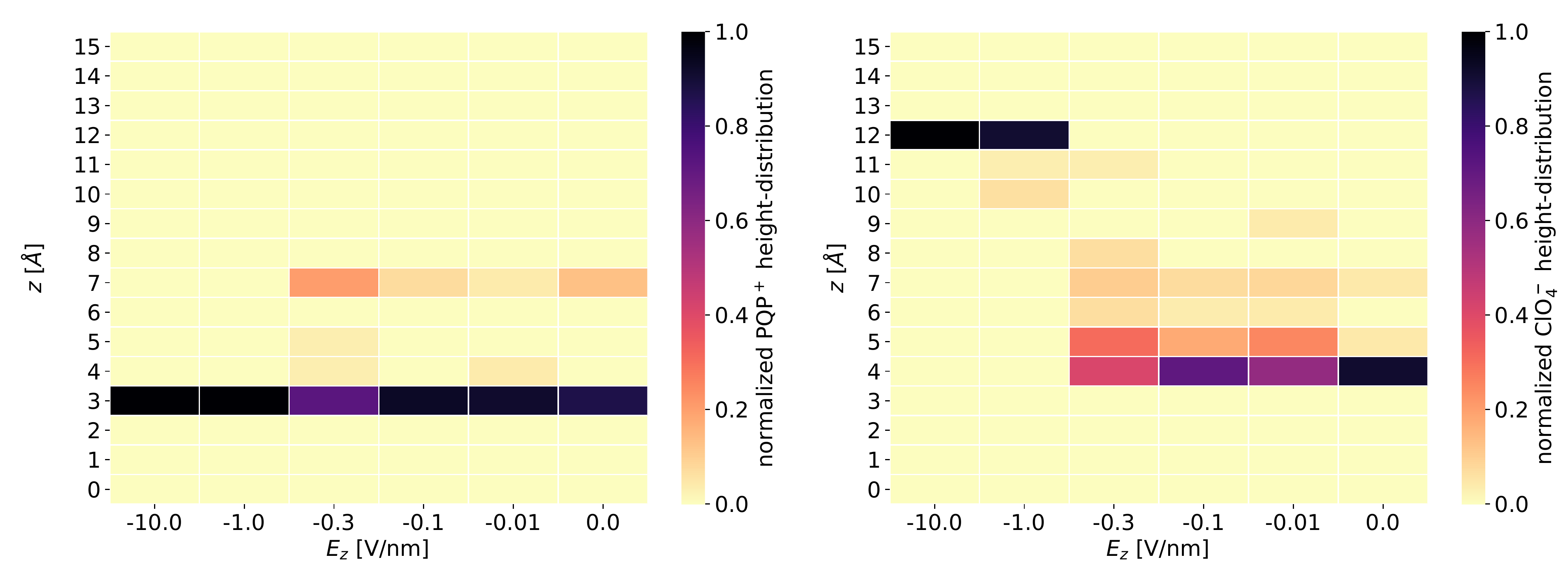}
\caption{(color online) Height-distribution of PQP$^+$
    (left) and ClO$_4^-$ (right) molecules as functions of the
    considered values of the electrostatic field, $E_z$, normalized to
    the number of respective molecules (see colour code at the right
    hand side of the panels; the value of one means, that all
    respective molecules in all considered configurations are counted
    in one specific bin), see text for the energy considered in this
    analysis).  Along the vertical axes the binning is performed in
    steps of one \AA: $z=0$\AA$\,$ marks the position of the gold
    surface, the slab width amounts to $12$\AA.}
\label{fig:ea:results_height_distribution}
\end{center}
\end{figure}

In Fig.~\ref{fig:ea:results_height_distribution} we
present in separate panels the height distributions of PQP$^+$ and
ClO$_4^-$ as functions of the electrostatic field, $E_z$ (which is
binned for the six different values of $E_z$ that were
investigated); along the vertical axis we count (for a given value
of $E_z$) the occurrence of the respective molecules in bins of one
\AA, considering all configurations identified by the evolutionary
algorithm, which are located within an interval of at most
1k$_\mathrm{B}$T (or $43$~meV) above the configuration with the best
energy, which are of the same order of magnitude as the values
presented in Table~\ref{tab:results} for different electric field
strengths, $E_z$; these distributions are generated  by counting, at
a given value of $E_z$, the vertical occurrence of the respective
molecules in bins of one \AA, and are normalized by the total number
of encountered structures. Note in this context that the distance of
the first layer of PQP$^+$ molecules can be directly estimated by the
vertical equilibrium position of carbon atoms,
$z_{\mathrm{C}}^{\mathrm{(eq)}}=3.166$\AA\,, obtained by minimizing
Eq.~(\ref{eq:LJ-wall:10-4-3}) for a single carbon atom with
$\sigma_\mathrm{wC}^{\mathrm{(Mie)}}=3.208$, taken from
Table~\ref{tab:fit:atomistic:wall}.

For the high values of the field strength (i.e. for
$E_z=-10$~V/nm and $-1$~V/nm) the PQP$^+$ ions are preferentially
adsorbed onto the gold surface as a closely packed monolayer (see
left panel of Fig.~\ref{fig:ea:results_height_distribution}), while
the perchlorate anions are strongly dissociated and assemble as far
from the gold surface as possible (corresponding in our
investigation to the numerical value of the slab height, which we
fixed to $12$~\AA) -- see right panel in
Fig.~\ref{fig:ea:results_height_distribution}.  This situation
represents an extreme case in the sense that neither the Coulomb nor
the short-range Mie interactions between ions and anions can
compensate for the strong negative surface potential.

Decreasing now the magnitude of the electrostatic field to
the more moderate value of $E_z\sim-0.3$~V/nm reveals the emergence
of a bilayer structure, formed by the PQP$^+$ molecules, with
pronounced peaks located at $z_{\mathrm{PQP}^+}^{(1)}\sim 3$\AA$\,$
and $z_{\mathrm{PQP}^+}^{(2)}\sim 7$\AA$\,$, with relative weights
of 72\% and 21\%, respectively. Such stratified bilayer
configurations (as depicted in panel (a) of
Fig.~\ref{fig:ea:results_experimental_trend}) are in competition
with structures similar to ones shown in panel (d) of
Fig.~\ref{fig:ea:results_experimental_trend}.  Note that in parallel
a far more complex height distribution of the perchlorate molecules
sets in as soon as the now moderate electrostatic field allows them
to proceed towards the interior of the slab: now, more than half of
the ClO$_4^-$ anions are located ``in between'' the PQP$^+$
``layers'', trying on one side to compensate the charges of one or
several PQP$^+$ ``partners'' in the slab region and ``filling
spatial holes'' wherever they can, on the other side.  A large
portion of perchlorate ions is even allowed to adsorb on the surface
at a distance of $z_{\mathrm{ClO}_4^-}^{(1)}\sim 4.074$\AA\,; note
that these COM positions above the interface are larger than the
minimal height of the PQP$^+$ cations due to two reasons: if one face
of the oxygen tetrahedron is oriented towards the interface (i.e.,
parallel to the gold surface), the COM of the ClO$_4^-$ ion is
increased by a value of $z_\mathrm{Cl}-z_\mathrm{O}\approx0.492$\AA\,
with respect to the oxygen atoms. These atoms themself have an
equilibrium distance to the surface of $z_\mathrm{O}^\mathrm{(eq)}
\approx 3.582$ (evaluated by minimizing Eq.~(\ref{eq:LJ-wall:10-4-3})
for a single oxygen atom with $\sigma_\mathrm{wO}^{\mathrm{(Mie)}}=
3.630$, cf. Table~\ref{tab:fit:atomistic:wall}), summing up to
the presented minimal COM distance of the adsorbed ClO$_4^-$
molecules from the interface.  Note that the height distribution
of the ClO$_4^-$ ions is now rather broad (see
Fig.~\ref{fig:ea:results_perchlorate_stacked_structures}), which
is definitely owed to their relatively smaller size and their
considerably higher mobility, as compared to their cationic
counterparts (see also discussion below); these features make a
conclusive interpretation of the roles of the ClO$_4^-$ ions in the
structure formation of the entire system rather difficult. Our
interpretation is that the perchlorate ions are -- due to their
small spatial extent and their high mobility -- able to compensate
for local charge mismatches and to act as spatial spacers between
the cations.

Decreasing further the magnitude of the electrostatic
field down to $E_z\sim-0.1$~V/nm and $E_z\sim-0.01$~V/nm provides
unambiguous evidence that the formation of the PQP$^+$ ions into
bilayer structures become energetically more and more unfavourable,
as the upper peak in the height distribution of the cations vanishes
gradually. Concomitantly, an increasing number of perchlorate
molecules approach the gold surface and are predominantly located
there; possibly they act as a space filler on the surface itself
while at the same time the small values of the electrostatic
field keep the PQP$^+$ molecules near to the surface. In this
context it should be noted that decreasing the electrostatic field
is equivalent to decreasing the surface potential; thus and in
combination with the adsorbed perchlorate molecules a transition
from an auto-host--guest to a porous structure is plausible.

Eventually, at zero electric field the system exclusively gains
energy from intramolecular interactions and adsorbtion on the gold
surface. Since the perchlorate molecules are rather spherical in
their shape they can efficiently adsorb onto the gold surface (in
an orientation explained above), while the PQP$^+$ molecules are
able to efficiently stack, especially without a guiding
electrostatic field.

\subsubsection{The role of the perchlorate anions}
\label{subsubsec:results_perchlorates}

\begin{figure}[htbp]
\begin{center}
\includegraphics[width=0.9\textwidth,
  clip=True]{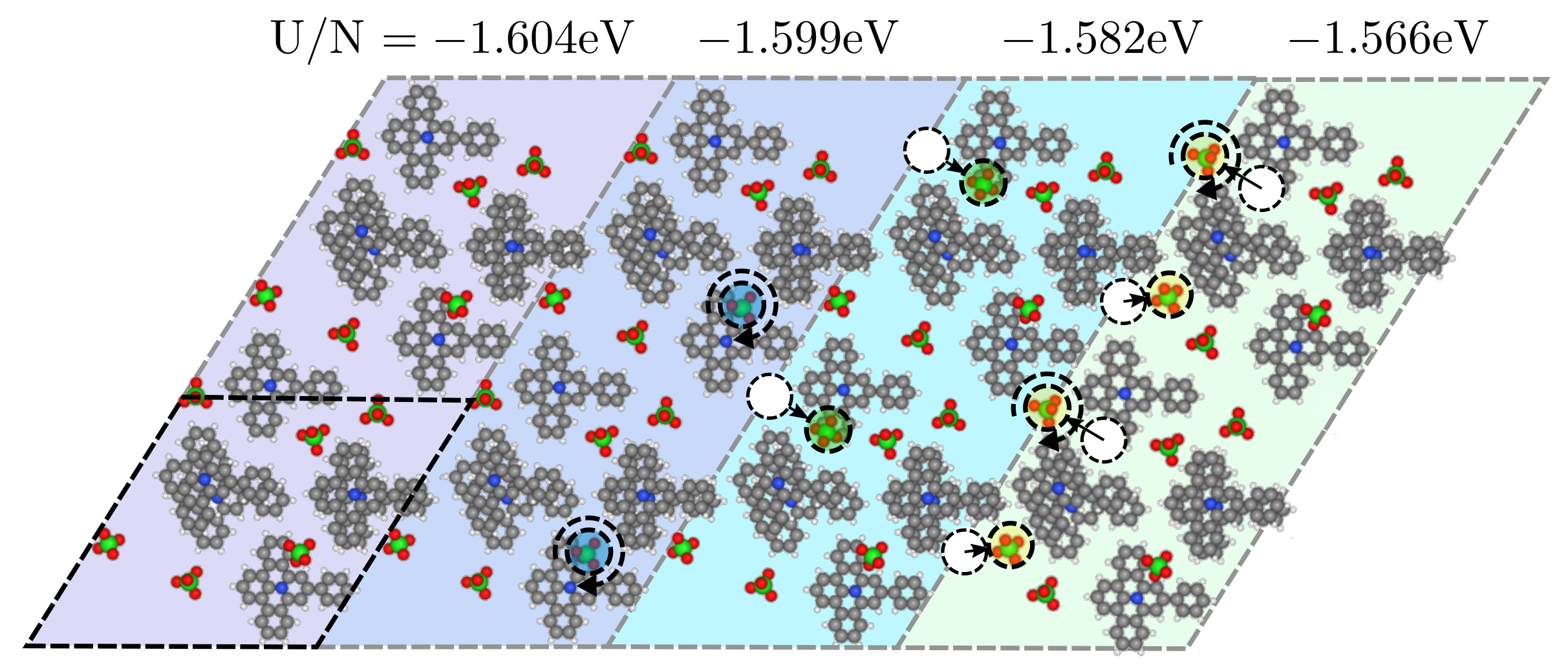}
\caption{(color online) Four structurally different
    configurations of perchlorate ions (framed by gray, dashed lines
    and differently shaded areas) at optimized, fixed cell geometry
    (indicated in the bottom left corner by the black, dashed line)
    and optimized, fixed positions and orientations of the PQP$^+$
    ions; the energies of these four configurations range within an
    interval of $38$~meV (per PQP$^+$--ClO$_4^-$ pair), as obtained
    after an evolutionary energy minimization of solely the degrees of
    freedom of the ClO$_4^-$ ions, starting from the configuration
    depicted in panel (a) of
    Fig.~\ref{fig:ea:results_experimental_trend}. Changes in the
    structure as one proceeds from left to right are highlighted by
    respective circles (specifying the position of the ``moving''
    perchlorate ion) and arrows.}
\label{fig:ea:results_perchlorate_stacked_structures}
\end{center}
\end{figure}

We come back to the above mentioned volatility of the
perchlorate ions: in
Fig.~\ref{fig:ea:results_perchlorate_stacked_structures} we present
results from yet another evolutionary analysis: we now fix the
positions and orientations of PQP$^+$ molecules as well as the
extent and the shape of the unit-cell of some optimized
configuration (as, for instance, depicted in panel (a) of
Fig.~\ref{fig:ea:results_experimental_trend}) and vary only the
degrees of freedom of the perchlorate anions.
Fig.~\ref{fig:ea:results_perchlorate_stacked_structures} shows --
for fixed cell geometry and fixed PQP$^+$ positions and orientations
-- four structurally different perchlorate arrangements whose energy
ranges within an interval of $38$~meV (per PQP$^+$--ClO$_4^-$ pair):
the fact that we obtain completely different configurations of the
perchlorates (with essentially comparable energies) undoubtedly
indicates the high mobility of the ClO$_4^-$ ions. Changes in the
structure, as one proceeds from left to right, are highlighted by
respective circles (specifying the position of the ``moving''
perchlorate ion) and related arrows.  The ClO$_4^-$ molecules
exhibit a remarkable freedom in their rotation without (or only
marginally) changing the energy of a configuration; this fact has
rendered the minimization of the energy very difficult. However, it
should also be noted that even translations can be performed without
a substantial change in energy.

We note that the analysis of these different structures was
achieved by using a so-called t-SNE\cite{Maaten2008} analysis on the
leading five PCA components\cite{Jolliffe2002} of order parameters
of all configurations identified by the evolutionary algorithm; for
more detailed information on this rather technical issue we refer to
Fig.~5 in subsection~4.2 of the Supplementary Information.

\section{Conclusion and outlook}
\label{sec:conclusion}

The prediction of supramolecular ordering of complex molecules at a
metal--electrolyte interface using DFT based {\it ab initio}
calculations is in view of the expected gigantic computational costs,
and despite the availability of peta-scale computers, still an elusive
enterprise.  In this contribution we have proposed a two-stage
alternative approach: (i) DFT-based {\it ab initio} simulations
provide reference data for the energies introduced in a classical
model for the molecules involved, where each of their atomic entities
are represented by a classical, spherical particle (with respective
size, energy parameters, and charges).  We modelled the interaction
between the atomic entities and the metallic surface by a classical,
perfectly conductive, Lennard-Jones like wall potential; the
electrolyte is treated as a homogeneous, dielectric medium.

The inter-particle and particle-wall parameters were obtained via the
following procedure: considering archetypical configurations
(involving pairs of ions and/or ions located close to the surface) DFT
energies were fitted by the related energy values of the classical
model.  (ii) The second step identifies the ordered ground state
configurations of the molecules by minimizing the total energy of the
now classical system.  This optimization is based on evolutionary
algorithms, which are known to operate efficiently and reliably even
in high dimensional search spaces and for rugged energy surfaces.

Our new two-stage strategy overcomes the hitherto prohibitive
computational cost of modelling the full system, while reproducing the
key observations of a well-documented experimental system consisting
of disc-shaped PQP$^+$ cations and ClO$_4^-$ anions: as a function of
increasing electric field at the metal--electrolyte interface, the
molecular building blocks are seen to self-organize into an open
porous structure, a self-host--guest pattern and a stratified bilayer.
Future work will focus on verifying the extent of predictive power of
our method towards molecular self-assembly under electrochemical
conditions, and on strategies to further streamline and reduce the
computational cost of our approach, without sacrificing the
reliability of the predicted results.

\begin{figure}
\includegraphics[width=\textwidth]{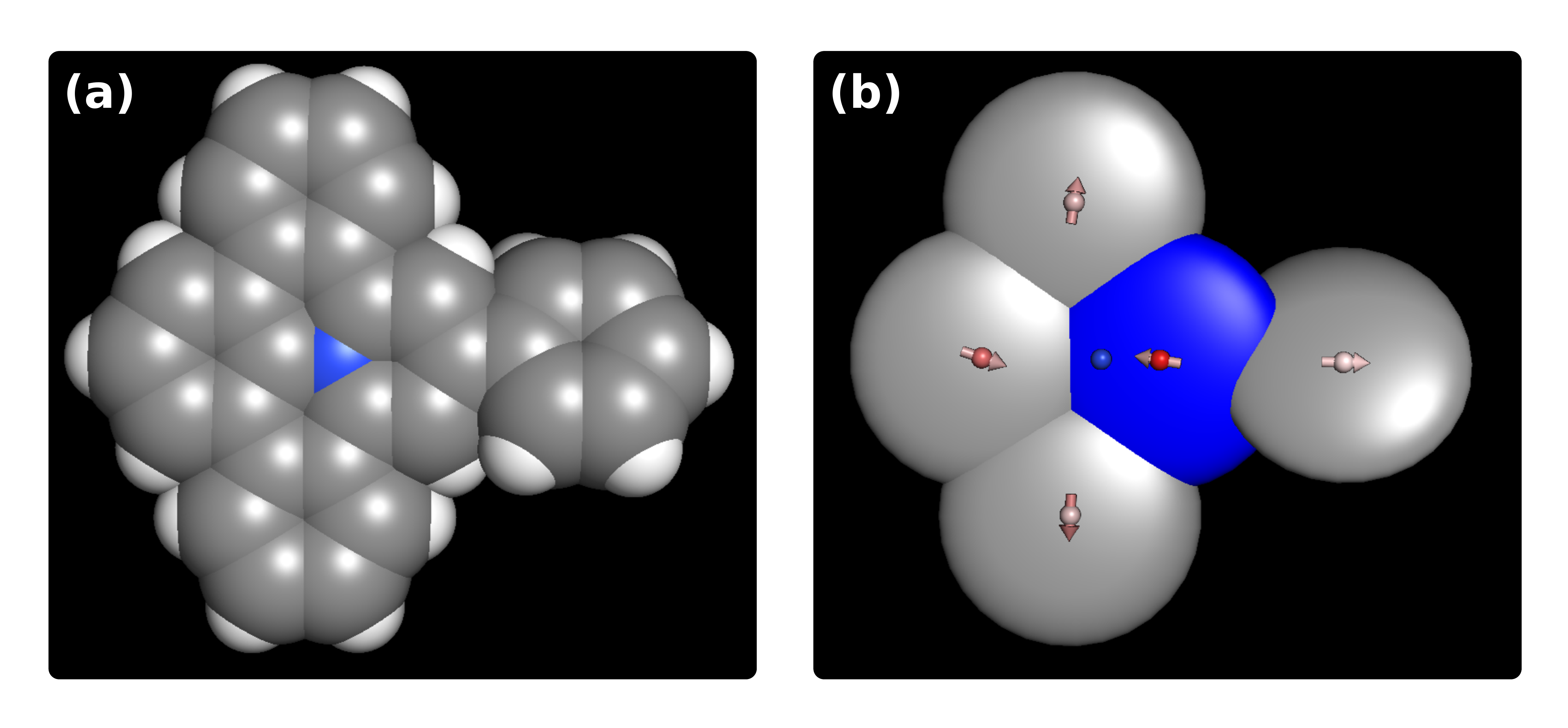}
\caption{Panel (a): atomistic model of the PQP$^+$
    molecule as used in this contribution (white: hydrogen atoms,
    grey: carbon atoms, and blue: nitrogen atom), see also
    Fig.~3 in the S.I.; panel (b): related
    coarse-grained model in a hierarchy of ever simpler models, using,
    e.g., Gay-Berne potentials to account for the van der Waals
    interaction of all atoms in the specific rings and a multi-pole
    expansion to second order (monopole as coloured points, dipole
    moments as small arrows) for the electrostatic interaction.}
  \label{fig:outlook:coarse}
\end{figure}

In view of the high computational costs and the conceptual
challenges encountered in our investigations we have pondered the
question if the complexity of the current model (which -- as a
classical model -- is comprehensive in the sense that it contains
all atomistic features) could possibly be further reduced, avoiding
thereby conceptual and computational bottlenecks. The idea behind
this strategy is to develop -- starting from the present model -- a
hierarchy of ever simpler models where, for instance, larger
sub-units of the molecule (such as aromatic rings) are replaced by
disk shaped units carrying higher electrostatic moments, as
schematically visualized in Fig.~\ref{fig:outlook:coarse}. Such a
model might provide a first, semi-quantitative prediction of the
self-assembly of the PQP$^+$ and of the ClO$_4^-$ ions at
considerably reduced costs and might help to pre-screen possibly
promising portions of the huge parameter space for subsequent
investigations of the full model. Efforts in this direction are
currently pursued.

\section{Acknowledgment}

BH acknowledges a DOC Fellowship of the Austrian Academy of Sciences;
further he gratefully acknowledges helpful discussions with Moritz
Antlanger (Wien), Benjamin Rotenberg (Paris), and Martial Mazars
(Paris-Sud) on several issues of this project and thanks the Freiburg
Centre for Interactive Materials and Bioinspired Technologies
(Universit\"at Freiburg) for the kind hospitality, where part of the
work has been carried out.  BH also acknowledges the Christiana
H\"orbiger prize for covering travel expenses to Paris.  BH and GK
acknowledge financial support by E-CAM, an e-infrastructure center of
excellence for software, training and consultancy in simulation and
modelling funded by the EU (Proj. No. 676531).  The computational
results presented have been achieved [in part] using the Vienna
Scientific Cluster (VSC).  SS and MW acknowledge funding from Deutsche
Forschungsgemeinschaft (WA 1687/10-1) and the computing time granted
by the John von Neumann Institute for Computing (NIC) within project
HFR08 as well as the computational resource bwUni-Cluster funded by
the Ministry of Science, Research and the Arts Baden-W\"urttemberg and
the Universities of the State of Baden-W\"urttemberg, Germany, within
the framework program bwHPC.  MW is thankful for enlightening
discussion with Johannes Fiedler and Stefan Buhmann.  SFLM and GK
gratefully acknowledge financial support within the TU Wien funded
Doctoral College BIOINTERFACE.  SFLM further acknowledges support from
the Austrian Science Fund (FWF, ``Boron Nitride Nanomesh for Actuated
Self-Assembly'' (I3256-N36)) and Energy Lancaster.

\begin{table}[htbp]
    \adjustbox{max width=\textwidth}{%
\begin{tabular}{
    p{0.12\textwidth}
    p{0.08\textwidth} p{0.08\textwidth} p{0.08\textwidth} p{0.08\textwidth} p{0.08\textwidth}
    p{0.08\textwidth}p{0.08\textwidth}p{0.08\textwidth}p{0.08\textwidth}p{0.08\textwidth}}
\toprule
{} &  $\sigma_{\rm H}$ &  $\sigma_{\rm C}$ &  $\sigma_{\rm N}$ &  $\sigma_{\rm O}$ &
$\sigma_{\rm Cl}$ &  $\epsilon_{\rm H}$ &  $\epsilon_{\rm C}$ &  $\epsilon_{\rm N}$ &
$\epsilon_{\rm O}$ &  $\epsilon_{\rm Cl}$ \\
\midrule
  $\mathcal{M}^{(\rm LJ)}$  & 2.243 & 3.658 & 3.743 & 2.865 & 5.953 & 3.052 & 1.204 & 3.311 & 7.396 & 0.172 \\  %
  $\mathcal{M}^{(\rm Mie)}$ & 2.236 & 3.703 & 3.328 & 2.428 & 4.956 & 3.999 & 0.946 & 2.021 & 11.481 & 5.289 \\  %
\midrule
\midrule
{} &  $\MieRep{\mathrm{H}}$ &  $\MieRep{\mathrm{C}}$ &  $\MieRep{\mathrm{N}}$ &  $\MieRep{\mathrm{O}}$ &
$\MieRep{\mathrm{Cl}}$ &  $\MieAtr{\mathrm{H}}$ &  $\MieAtr{\mathrm{C}}$ &  $\MieAtr{\mathrm{N}}$ &
$\MieAtr{\mathrm{O}}$ &  $\MieAtr{\mathrm{Cl}}$ \\
\midrule
  $\mathcal{M}^{(\mathrm{LJ})}$  &   12 &   12 &   12 &   12 &   12 &   6 &   6 &   6 &   6 &    6 \\ %
  $\mathcal{M}^{(\mathrm{Mie})}$ & 6.263 & 7.136 & 8.659 & 8.743 & 15.455 & 7.500 & 12.299 & 13.854 & 17.193 & 4.684 \\ %
\bottomrule
\end{tabular}
    }
\caption{
  Numerical results for the optimized LJ and Mie parameters,
  $\mathcal{M}^{(\mathrm{LJ})}=\{\sigma_i, \epsilon_i\}$ and
  $\mathcal{M}^{(\mathrm{Mie})}=\{\sigma_i, \epsilon_i, \MieRep{i}, \MieAtr{i}\}$,
  for each element $i$, for the results depicted in the Fig.~\ref{fig:fit:atomistic}
  ($\sigma_i$ in \AA\ and $\epsilon_i$ in meV).
  Reference values from the literature are listed in
  Table~1 in the supplementary information.
}
\label{tab:fit:atomistic:vdw}
\end{table}

\begin{table}[htbp]
    \adjustbox{max width=\textwidth}{%
\begin{tabular}{p{0.1\textwidth}rrrr}
\toprule
{} & $\sigma_\mathrm{ w [H, C, N]}$ &  $\sigma_\mathrm{ w [O, Cl]}$ &
$\epsilon_\mathrm{w [H, C, N]}$ &  $\epsilon_\mathrm{w [O, Cl]}$ \\
\midrule
  {$\mathcal{W}^{(\mathrm{LJ})}$}  & 3.197 & 3.625 & 3.741 & 15.781 \\  %
  {$\mathcal{W}^{(\mathrm{Mie})}$} & 3.208 & 3.630 & 3.698 & 20.167 \\  %
\bottomrule
\end{tabular}
    }
\caption{ Top row: Numerical results for LJ-length and well-depth parameters,
  $\sigma_{{\mathrm w}i}$ in \AA\ and $\epsilon_{{\mathrm w}i}$ in meV,
  between the wall and each element $i={\rm[H, C, N]}$ and $j={\rm [O, Cl]}$,
  grouped by the molecules they belong to (PQP$^+$ and ClO$_4^-$),
  for intermolecular short-range LJ parameters listed in Table.~\ref{tab:fit:atomistic:vdw}.
  Bottom row: corresponding $\sigma_{{\mathrm w}i}$ and $\epsilon_{{\mathrm w}i}$ parameters
  for intermolecular short-range Mie-parameters also listed in Table.~\ref{tab:fit:atomistic:vdw}.
}
\label{tab:fit:atomistic:wall}
\end{table}

\begin{table}[htbp]
    \adjustbox{max width=\textwidth}{%
\begin{tabular}{p{0.08\textwidth}lllll}   %
\toprule
 & $E_z$ & $U/N_{\rm PQP}$ & $N_{{\rm PQP}}$ & $A$ & $N_{\rm PQP}/A$  \\  %
 & [Vnm$^{-1}$] & [eV] &  &  [nm$^2$] & [nm$^{-2}$]                         \\  %
\midrule
(a) &   -0.30	    & -1.5804 	       & 6   & 8.5 	       & 0.705882 \\  %
(b) &   -0.10	    & -1.7276 	       & 6   & 11.75 	   & 0.510638 \\  %
(d) &   -0.10	    & -1.7274	       & 5   & 8.0 	       & 0.625000 \\  %
(c) &   -0.01       & -1.6445          & 6   & 11.25 	   & 0.533333 \\  %
(f) &   -0.01       & -1.6364 	       & 6   & 12.25 	   & 0.489796 \\  %
(e) &   -0.01       & -1.6332          & 7   &	11.75 	   & 0.595745 \\  %
\bottomrule
\end{tabular}
    }
\caption{Results of evolutionary ground-state search for different
  electric field strengths, $E_z$, for different unit-cell areas, $A$ and number
  of ${\rm PQP}^+$ molecules, $N_{{\rm PQP}}$, each line represents a evolutionary search.
  The respective structures are presented in
  Fig.~\ref{fig:ea:results_experimental_trend}.
}
\label{tab:results}
\end{table}

\bibliography{main}

\end{document}


\maketitle

\section{Introduction}
\label{app:introduction}

In the Supplementary Information (S.I.) we have collected relevant
information which might considerably deteriorate the readability of
the main text if it were placed there; still, the details presented in
this document might be of relevance for an interested reader of the
main text. For simplicity we have used in this document exactly the
same section headings as in the main text; this will hopefully help to
establish the appropriate association between the respective text
passages.

\section{The system and its representations}
\label{app:system}

\subsection{Convergence of theoretical calculations}

\begin{figure}[htbp]
  \subfloat[BE = -2.69 eV]{{\includegraphics[width=0.31\linewidth]{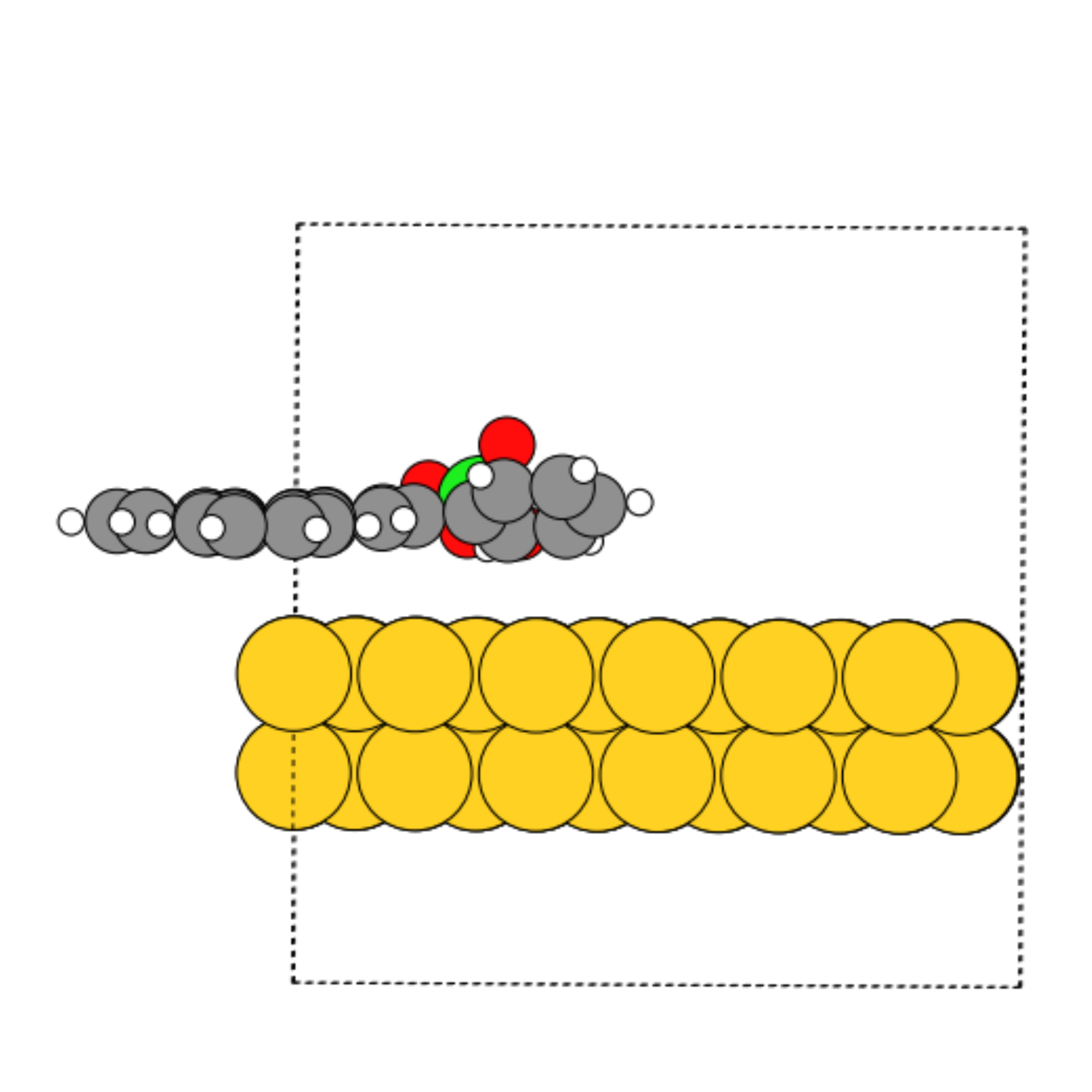} }}%
  \hspace{0.01\linewidth}
  \subfloat[BE = -2.66 eV]{{\includegraphics[width=0.31\linewidth]{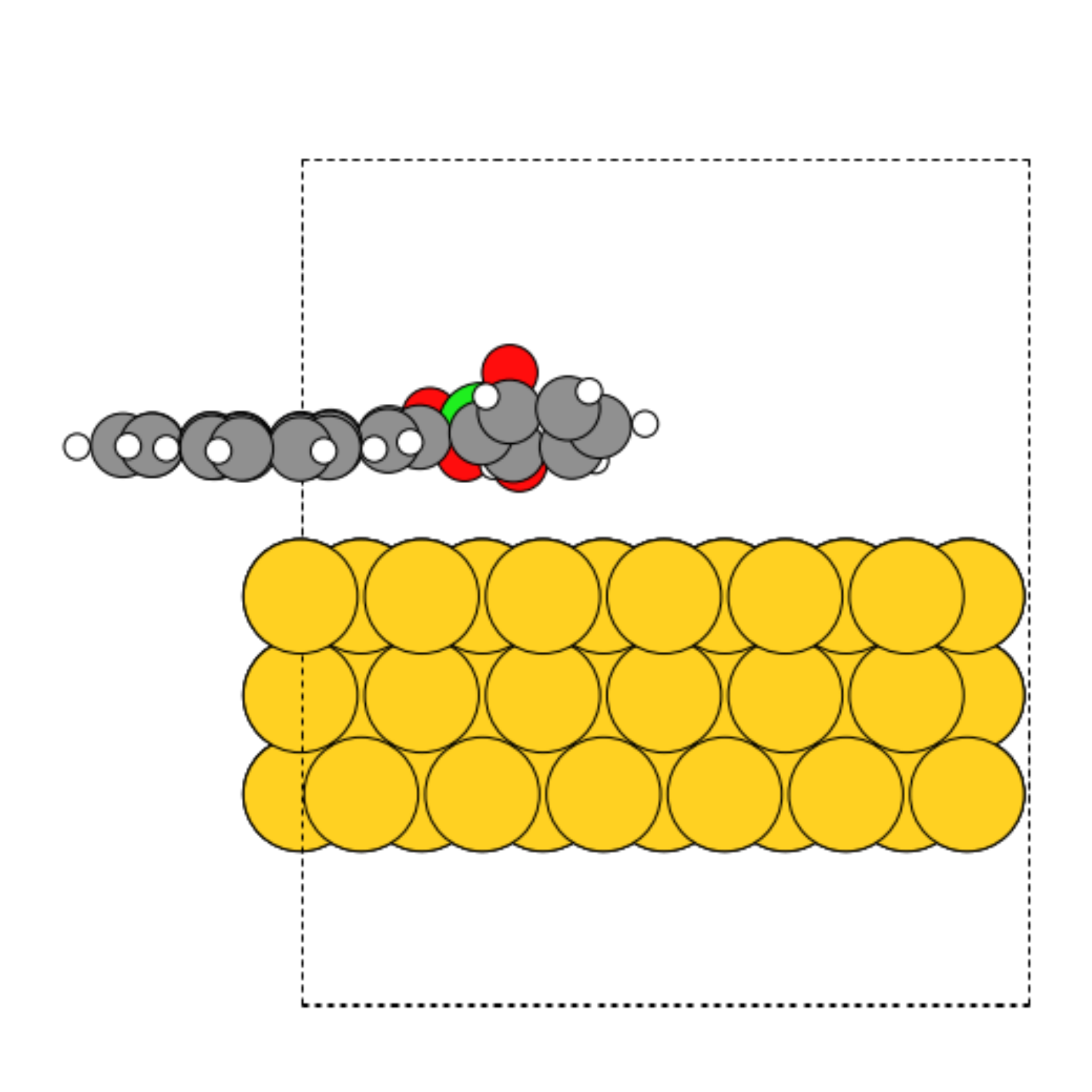} }}%
  \hspace{0.01\linewidth}
  \subfloat[BE = -2.71 eV]{{\includegraphics[width=0.31\linewidth]{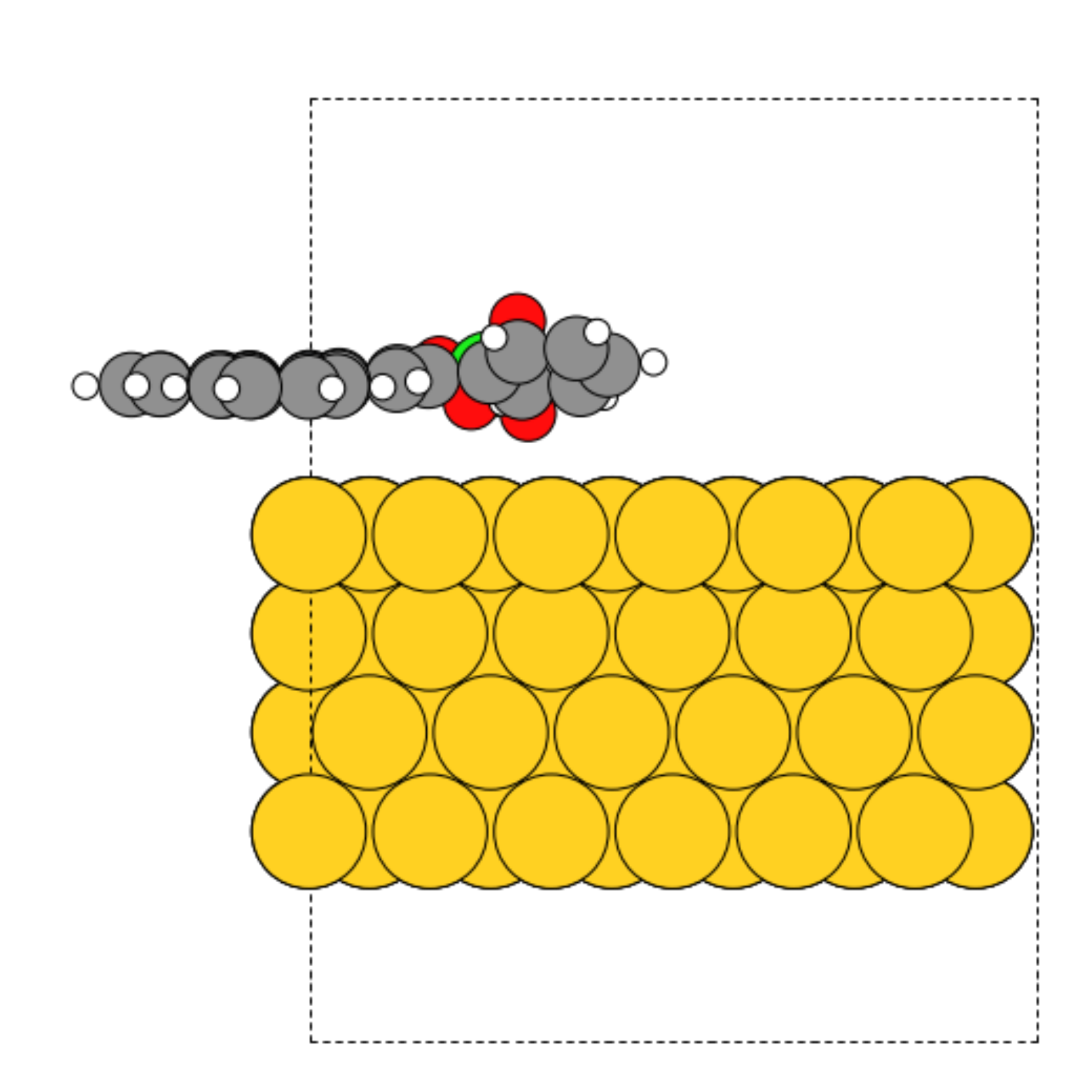} }}%
  \caption{Schematic views of PQP$^+$ and ClO$_4^-$ ions, located
    above gold surfaces built up by different numbers of layers Below
    each panel the respective binding energies (BE) are specified.  }
  \label{fig:fit:BE for different gold layers}
\end{figure}

The DFT-based binding energies of the PQP$^+$ and ClO$_4^-$ ions were
calculated for different numbers of gold layers with a
  constraint of fixed gold layers in an effort to study the
convergence of the results with respect to the number of layers that
build up the gold surface. The values for the binding energies,
obtained for the different cases, are specified below the respective
panels of Fig.~\ref{fig:fit:BE for different gold layers}; the panels
themselves provide schematic plots for the different types of gold
layers. In the related DFT calculations we did not consider the van
der Waals scaling, i.e., we chose $\omega_S\equiv 1$ in
Eq.~(1) of the main text.
\vspace{0.2cm}

 We further examined the effect of fixing the gold
atoms on the binding energy of PQP$^+$ and ClO$_4^-$ ions to the
surface. Allowing the uppermost gold layer to relax changed the
binding energies to -2.73 eV, -2.69 eV and -2.73 eV for two, three
and four layers, respectively.  These values are well in accordance
with the previously calculated values reported in
Fig.~\ref{fig:fit:BE for different gold layers}.  We therefore can
conclude that fixing the gold atoms has a negligible effect on the
binding energy obtained.

The calculated values for the binding energy obtained for two, three,
and four layers of gold provide evidence that our choice for a two
layer gold surface is sufficient to proceed with our calculation of
the self-assembly scenarios of PQP$^+$ and ClO$_4^-$ ions on this
Au(111) surface. Local density of states were also plotted with
respect to the energies of the corresponding states relative to the
Fermi-level.  The plotted density curves shows that the density of
states for PQP$^+$ and ClO$_4^-$ ions are quite similar irrespective
of the number of layers of gold surface. \\

\begin{figure}[htbp]
  \subfloat[two layers]{\includegraphics[width=0.32\linewidth]{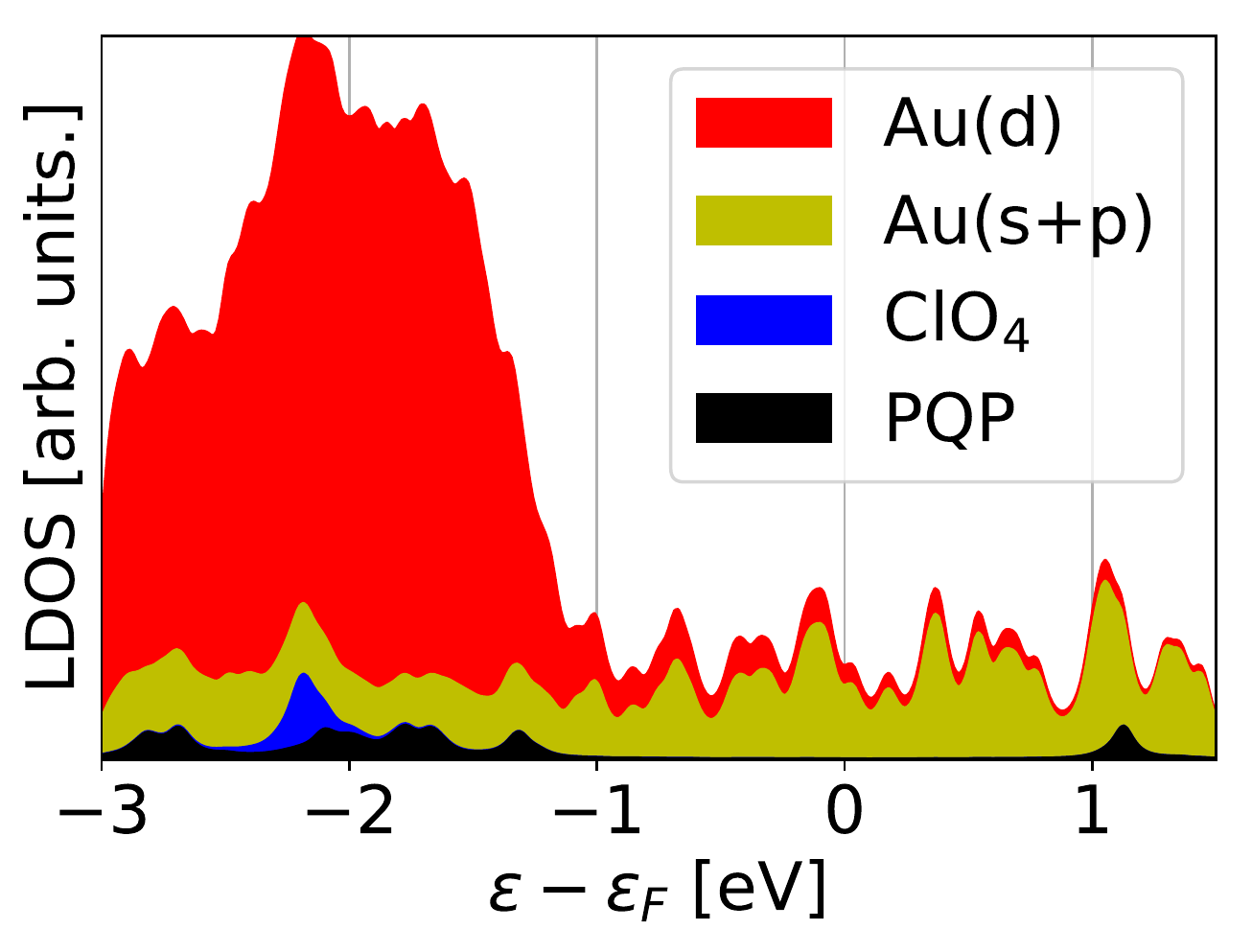}}
  \hspace{0.01\linewidth}
  \subfloat[three layers]{\includegraphics[width=0.32\linewidth]{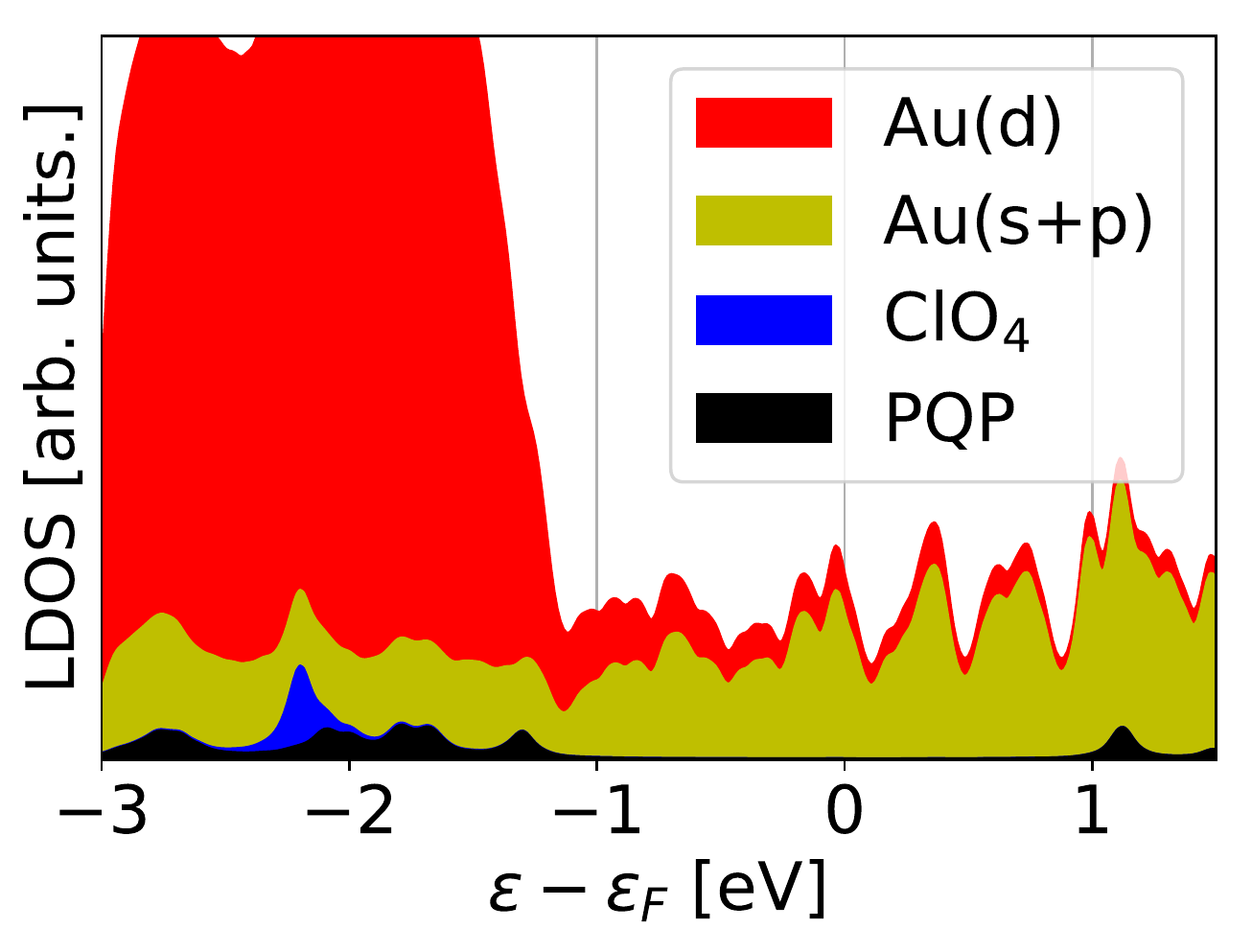}}
  \hspace{0.01\linewidth}
  \subfloat[four layers]{\includegraphics[width=0.32\linewidth]{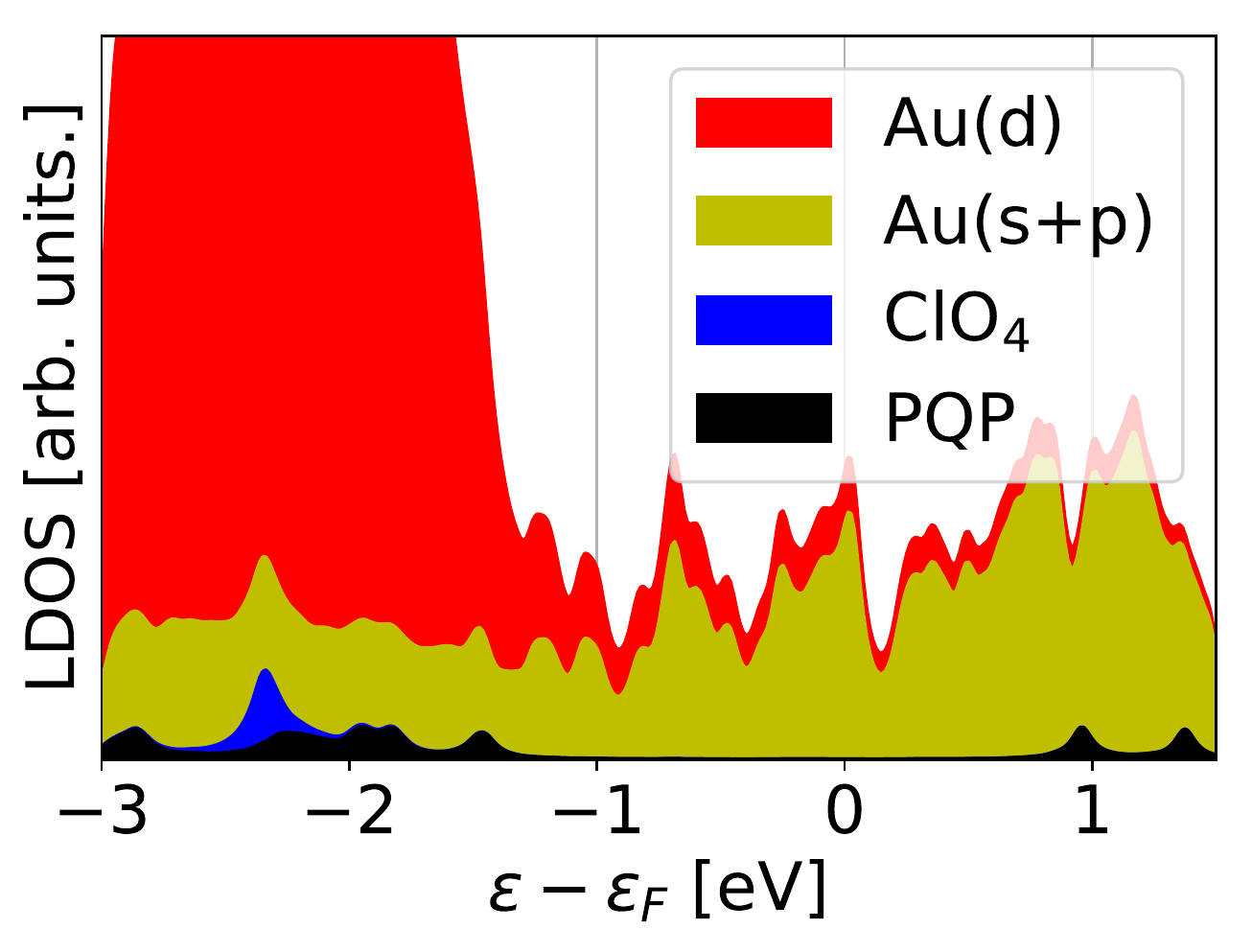}}
  \caption{Local density of states (LDOS) relative to
      Fermi-level.  The Kohn-Sham energies were projected on local
      atomic orbitals of Cl and O (ClO$_4$), C, N and H (PQP), and on
      gold s+p and d states, respectively.}
  \label{fig:fit:BE for different gold layers:ldos}
\end{figure}

Fig.~\ref{fig:fit:BE for different gold layers:ldos} compares the
density of Kohn-Sham states projected on the atomic species for the
models depicted in Fig.~\ref{fig:fit:BE for different gold layers}.
While the density of states of the gold part increases naturally with
the number of gold layers, this has only marginal effects on the
positions of the PQP$^+$ and ClO$_4^-$ related states.

\subsection{Angle-axis framework expressing rigid body orientations}
\label{app:axis_angle_framework}

In the optimization procedure put forward in
Section~3 of the main text we rely on the
angle-axis framework \cite{Chakrabarti2008, Chakrabarti2014} to
express the orientation of rigid molecules within the lab-frame:
closely related to the descriptions of rotations based on
unit-quaternions \cite{Evans1997,Allen1989,Miller2002}, we introduce a
three-component angle-axis vector, $\MoleculeOrientation{} =
\MoleculeOrientationVector=\theta\,\hat{\MoleculeOrientation{}}$,
which defines an angle, $\theta=\MoleculeOrientationAngle$, and a
unit-vector, $\hat{\MoleculeOrientation{}}$, which represents the axis
of the molecule; both are sufficient to describe any rotation of a
rigid body in three dimensions.  As discussed in
Ref.~\citenum{Chakrabarti2014} and following Rodrigues' rotation
formula, the $3\times3$ rotation matrix
$\RotationMatrix{\MoleculeOrientation{}}$ associated with the
angle-axis vector $\MoleculeOrientation{}$ is given by

\begin{equation}
  \RotationMatrix{\MoleculeOrientation{}} = \mathbb{I} +
  \left(1 - \cos{\theta}\right)
  \Skew{P}\Skew{P} +
  \left(\sin{\theta}\right)\Skew{P} ;
  \label{eq:rotation:rodrigues}
\end{equation}
here $\mathbb{I}$ is the $3\times3$ identity matrix and
$\Skew{P}$ is the skew-symmetric $3\times3$ matrix obtained
from the components of the vector $\hat{\MoleculeOrientation{}}$
via

\begin{equation}
  \Skew{P} = \dfrac{1}{\theta}\begin{pmatrix}
       0 & -P_3 &  P_2 \\
     P_3 &    0 & -P_1 \\
    -P_2 &  P_1 &    0 \\
  \end{pmatrix}.
  \label{eq:rotation:skew}
\end{equation}

In order to transform the coordinates of an atom,
$\CoordinateVector{m}{I}$, defined in the center-of-mass system of
molecule $\Molecule{I}$ to its lab-frame position,
$\PositionVector{m}$, the following transformation needs to be
realized:

\begin{equation}
  \PositionVector{m} = \MoleculePosition{I} +
\RotationMatrix{\MoleculeOrientation{I}}\cdot\CoordinateVector{m}{I} ;
\end{equation}
here $\MoleculePosition{I}$ is the center-of-mass coordinate of
molecule $\Molecule{I}$ and $\RotationMatrix{\MoleculeOrientation{I}}$
is the rotation matrix associated with the angle-axis vector
$\MoleculeOrientation{I}$ of molecule $\Molecule{I}$, as defined
above.

\subsection{Short-range potentials and parametrization}
\label{app:mie_potential:parametrization}

The short-range Mie potential, defined in Eq.~(5) of the
main text (Subsection~2.3.1), can be
considered as a generalization of the Lennard-Jones (LJ) interaction
\cite{Mie1903}: if the exponents of the repulsive and attractive parts
of the potential are chosen as $\MieRep{ij}=12$ and $\MieAtr{ij}=6$
the amplitude $C_{ij}$, given by Eq.~(6) of the main
text, becomes $C_{ij}=4$ and the Mie-potential reduces to the well
known LJ form.

During the fitting procedure put forward in
Subsection~2.3.2 of the main text, it occurred
at some instances that $\MieRep{i} < \MieAtr{i}$. In such a case the
defining equation for the $C_{ij}$ (see Eq.~(5) of the main
text) guarantees that both the repulsive and the attractive parts of
the potential maintain their respective features.

\begin{figure}[htbp]
\vspace{1.5cm}
  \includegraphics[width=0.7\textwidth]{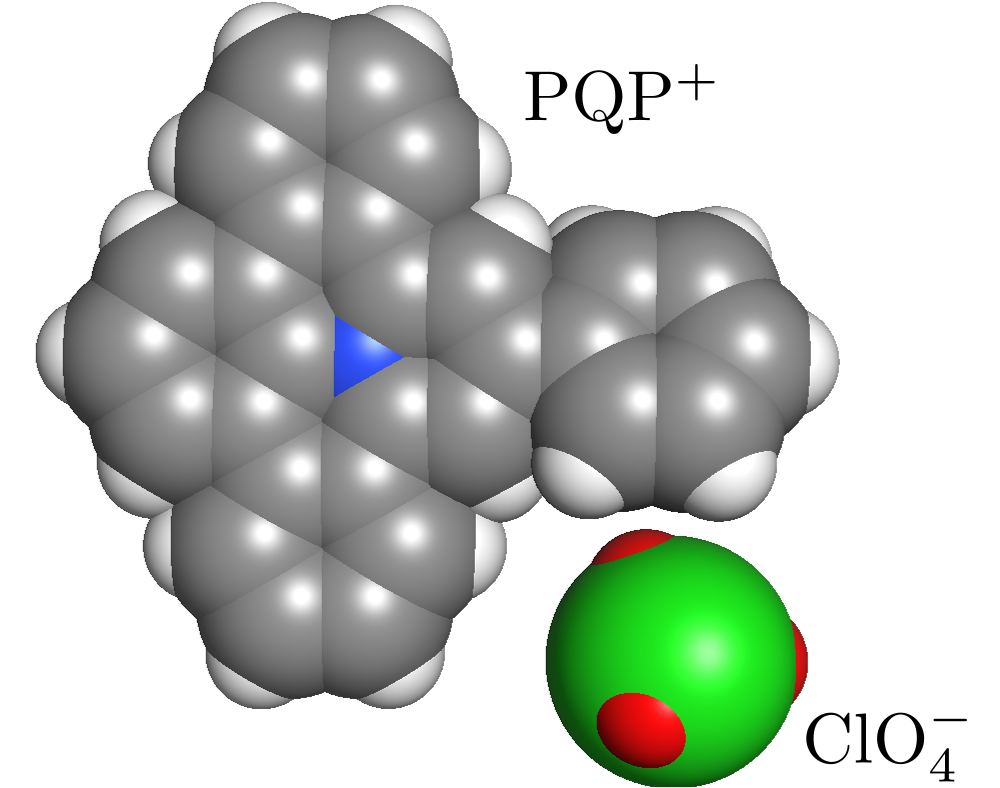}
 \vspace{0.5cm}
  \caption{(Color online) Schematic representation of a PQP$^+$ and a
    ClO$_4^-$ ion using the actual values of the fitted Mie length
    parameters $\sigma_i^{\rm (Mie)}$ for the short-range interactions
    introduced in Subsection~2.3.1 of the
    main text: atomic entities are shown as spheres with their
    diameters fixed by their respective optimized $\sigma_i^{\rm
      (Mie)}$-values; these entities are colored according to the
    following scheme: hydrogen (white), carbon (gray), nitrogen
    (blue), chlorine (green), and oxygen (red).  }
  \label{fig:app:Mie}
\end{figure}

In Fig.~\ref{fig:app:Mie} we depict the PQP$^+$ and the ClO$_4^-$ ions
using the actual values for the fitted Mie length parameters,
$\sigma_i$ (listed in Table~1 of the main
text), as van der Waals radii. In
Table~\ref{tab:fit:atomistic:vdw:reference} of the S.I. we list -- for
comparison -- also the LJ length parameters for the atomic entities of
the PQP$^+$ and the ClO$_4^-$ molecules as they are commonly used in
literature.

\begin{table}[htbp]
    \adjustbox{max width=\textwidth}{%
  \begin{tabular}{p{0.2\textwidth} p{0.16\textwidth}p{0.16\textwidth}p{0.16\textwidth}p{0.16\textwidth}p{0.16\textwidth}}
  \toprule
      {} &  $\sigma_{\rm H}$ &  $\sigma_{\rm C}$ &  $\sigma_{\rm N}$ &
      $\sigma_{\rm O}$ &  $\sigma_{\rm Cl}$ \\
  \midrule
  {\textnormal {mendeleev}} &   2.2   &   3.4    &   3.1    &   3.04   &    3.5    \\
  {\textnormal {alvarez}}   &   2.4   &   3.54   &   3.32   &   3.0    &    3.64   \\
  {\textnormal {bondi}}     &   2.4   &   3.4    &   3.1    &   3.04   &    3.5    \\
  {\textnormal {dreiding}}  &   3.195 &   3.8983 &   3.6621 &   3.4046 &    3.9503 \\
  {\textnormal {mm3}}       &   3.24  &   4.08   &   3.86   &   3.64   &    4.14   \\
  {\textnormal {uff}}       &   2.886 &   3.851  &   3.66   &   3.5    &    3.947  \\
  {$\sigma_i^{\rm(LJ)}$} &  2.243  &  3.658   &   3.743   &   2.865    &  5.953  \\
  {$\sigma_i^{\rm(Mie)}$} &  2.236  &  3.703   &  3.328   &   2.428    &  4.956  \\
  \bottomrule
  \end{tabular}
    }
  \caption{LJ length parameters $\sigma_i$ (in \AA) for hydrogen (H),
    carbon (C), nitrogen (N), oxygen (O), and chlorine (Cl) from
    literature as defined in the {\it mendeleev}-python-module ({\it
      ver. 0.5.1}) and, as comparison, model parameters
    $\sigma_i^{\rm(LJ)}$ and $\sigma_i^{\rm(Mie)}$ from
    Table~1 in the main text.  }
  \label{tab:fit:atomistic:vdw:reference}
\end{table}
\FloatBarrier

\begin{figure}[htbp]
  \begin{center}
    \includegraphics[width=0.7\textwidth, clip]{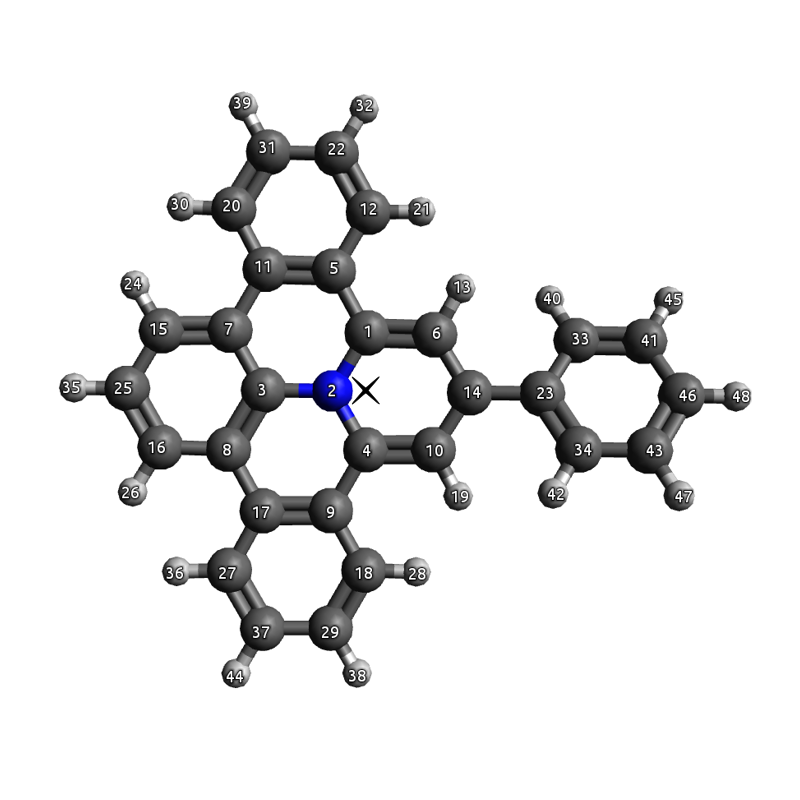}
    \caption{(Color online) Schematic view of the PQP$^+$ ion where
      its atomic constituents and the related bonds are depicted. The
      spheres are colored according to the respective chemical
      element: hydrogen (white), carbon (gray), and nitrogen (blue).
      The atomic constituents are labeled by indices $i=1$ to $N_{\rm
        PQP}=48$; the positions of each of these entities (with
      respect to the center-of-mass of the molecule) are listed in
      Table~\ref{tbl:pqp:xyzq}.  The black cross marks the center of
      mass, $\MoleculePosition{\rm PQP}$, of the PQP$^+$ molecule.  }
    \label{fig:app:pqp:indices}
  \end{center}
\end{figure}

In Tables~\ref{tbl:pqp:xyzq} and \ref{tbl:perchlorate:xyzq} we list
the coordinates of all atomic units and their associated partial
charges (extracted via a Bader analysis \cite{Bader1985,Tang09jpc})
obtained from the relaxed DFT-structures of the PQP$^+$ and ClO$_4^-$
ions which serve as rigid molecular blueprints in the main text.
Since the PQP$^+$ cation is built up by 48 atomic units we have
supplemented Table~\ref{tbl:pqp:xyzq} by
Fig.~\ref{fig:app:pqp:indices}, indicating the labeling of the
different atomic entities. In contrast, as the ClO$_4^-$ molecule (see
Table~\ref{tbl:perchlorate:xyzq}) is only built up by five atomic
units, we have refrained in this case from a schematic presentation of
the molecule.

\begin{table}[htbp]
  \adjustbox{max width=\textwidth}{%
\begin{tabular}{ccccccccccccc}
\toprule
{$i$} & element &  $x_i$ &  $y_i$ &  $z_i$ &  $q_i$ ~ & {~~~~~~~~$i$} & element &  $x_i$ &  $y_i$ &  $z_i$ &  $q_i$ \\
\midrule
1~  &       C &    0.1267 &    1.2051 &   -0.0253 &   0.3807 ~ & ~~~~~~~~25~ &       C &   -4.8220 &    0.0601 &   -0.0341 &  -0.0565 \\
2~  &       N &   -0.5941 &    0.0077 &   -0.0830 &  -1.1646 ~ & ~~~~~~~~26~ &       H &   -4.7122 &   -2.0727 &   -0.0499 &   0.1101 \\
3~  &       C &   -2.0123 &    0.0254 &   -0.0936 &   0.3949 ~ & ~~~~~~~~27~ &       C &   -2.7268 &   -3.7034 &   -0.1427 &  -0.0931 \\
4~  &       C &    0.0988 &   -1.2077 &   -0.0664 &   0.4209 ~ & ~~~~~~~~28~ &       H &    1.1435 &   -3.7411 &   -0.1704 &   0.1265 \\
5~  &       C &   -0.5685 &    2.4836 &   -0.0704 &   0.0642 ~ & ~~~~~~~~29~ &       C &   -0.6337 &   -4.9026 &   -0.1737 &  -0.1096 \\
6~  &       C &    1.5102 &    1.1636 &    0.0881 &  -0.0066 ~ & ~~~~~~~~30~ &       H &   -3.7189 &    3.8240 &   -0.1608 &   0.0990 \\
7~  &       C &   -2.7073 &    1.2652 &   -0.0766 &   0.0497 ~ & ~~~~~~~~31~ &       C &   -1.9108 &    4.9519 &   -0.1507 &  -0.0561 \\
8~  &       C &   -2.7389 &   -1.1964 &   -0.0946 &   0.0300 ~ & ~~~~~~~~32~ &       H &    0.0658 &    5.8448 &   -0.1588 &   0.0863 \\
9~  &       C &   -0.6299 &   -2.4673 &   -0.1204 &  -0.0401 ~ & ~~~~~~~~33~ &       C &    4.4671 &    1.0567 &   -0.2093 &  -0.0996 \\
10~  &       C &    1.4854 &   -1.2014 &    0.0175 &  -0.0272 ~ & ~~~~~~~~34~ &       C &    4.3714 &   -1.1639 &    0.7541 &  -0.0342 \\
11~ &       C &   -1.9791 &    2.5216 &   -0.0929 &  -0.0412 ~ & ~~~~~~~~35~ &       H &   -5.9114 &    0.0745 &   -0.0008 &   0.1122 \\
12~ &       C &    0.1514 &    3.7017 &   -0.0977 &  -0.0507 ~ & ~~~~~~~~36~ &       H &   -3.8146 &   -3.7276 &   -0.1477 &   0.0984 \\
13~ &       H &    2.0403 &    2.1040 &    0.1899 &   0.1396 ~ & ~~~~~~~~37~ &       C &   -2.0369 &   -4.9037 &   -0.1638 &   0.0333 \\
14~ &       C &    2.2322 &   -0.0280 &    0.1088 &   0.0058 ~ & ~~~~~~~~38~ &       H &   -0.0809 &   -5.8419 &   -0.2001 &   0.1258 \\
15~ &       C &   -4.1099 &    1.2499 &   -0.0437 &  -0.0944 ~ & ~~~~~~~~39~ &       H &   -2.4371 &    5.9066 &   -0.1831 &   0.1140 \\
16~ &       C &   -4.1411 &   -1.1474 &   -0.0630 &  -0.0382 ~ & ~~~~~~~~40~ &       H &    3.9741 &    1.9279 &   -0.6429 &   0.0888 \\
17~ &       C &   -2.0415 &   -2.4705 &   -0.1202 &   0.0066 ~ & ~~~~~~~~41~ &       C &    5.8578 &    1.0346 &   -0.1202 &  -0.0392 \\
18~ &       C &    0.0574 &   -3.7037 &   -0.1539 &  -0.0731 ~ & ~~~~~~~~42~ &       H &    3.8020 &   -2.0184 &    1.1245 &   0.0943 \\
19~ &       H &    1.9993 &   -2.1557 &    0.0110 &   0.1087 ~ & ~~~~~~~~43~ &       C &    5.7627 &   -1.1824 &    0.8408 &  -0.1172 \\
20~ &       C &   -2.6331 &    3.7708 &   -0.1343 &  -0.0780 ~ & ~~~~~~~~44~ &       H &   -2.5876 &   -5.8452 &   -0.1777 &   0.1138 \\
21~ &       H &    1.2385 &    3.7123 &   -0.0996 &   0.0676 ~ & ~~~~~~~~45~ &       H &    6.4355 &    1.8929 &   -0.4673 &   0.1010 \\
22~ &       C &   -0.5084 &    4.9176 &   -0.1347 &   0.0077 ~ & ~~~~~~~~46~ &       C &    6.5117 &   -0.0860 &    0.4007 &  -0.0581 \\
23~ &       C &    3.7024 &   -0.0450 &    0.2216 &  -0.0045 ~ & ~~~~~~~~47~ &       H &    6.2672 &   -2.0562 &    1.2563 &   0.1042 \\
24~ &       H &   -4.6584 &    2.1881 &   -0.0183 &   0.0912 ~ & ~~~~~~~~48~ &       H &    7.6004 &   -0.1048 &    0.4614 &   0.1070 \\
\bottomrule
\end{tabular}
  }
\caption{Atomic units building up the PQP$^+$ ion, labeled by the
  index $i$ according to the schematic view of the molecule presented
  in Fig.~\ref{fig:app:pqp:indices}. The positions of these entities
  ($x_i$, $y_i$ and $z_i$, and all in \AA), as they were obtained in a
  relaxed DFT-based configuration, are given with respect to the
  center-of-mass of the molecule, marked in this figure by a cross.
  Furthermore, the respective charges of the atomic units, $q_i$
  (given in units of the elementary charge $e$), are obtained in a
  Bader analysis \cite{Bader1985}; these charges are directly
  transferred to our classical model of the PQP$^+$ molecule.}
\label{tbl:pqp:xyzq}
\end{table}

\begin{table}[htbp]
  \adjustbox{max width=\textwidth}{%
\begin{tabular}{cccccc}
\toprule
$i$ & element & $x_i$ & $y_i$ & $z_i$ &  $q$ \\
\midrule
1~ &      Cl &    0.0000 &    0.0000 &    0.0000 &   2.6996 \\
2~ &       O &    1.4732 &   -0.0020 &    0.0000 &  -0.9249 \\
3~ &       O &   -0.4916 &    1.3888 &    0.0000 &  -0.9249 \\
4~ &       O &   -0.4917 &   -0.6933 &    1.2034 &  -0.9249 \\
5~ &       O &   -0.4917 &   -0.6933 &   -1.2034 &  -0.9249 \\
\bottomrule
\end{tabular}
  }
\caption{Atomic units building up the ClO$_4^-$ molecule. The
  positions of these entities ($x_i$, $y_i$ and $z_i$, and all in
  \AA), as they were obtained in a relaxed DFT-based configuration,
  are given with respect to the center-of-mass of the molecule, which
  coincides with the position of the oxygen atom.  Furthermore, the
  respective charges of the atomic units, $q_i$ (given in units of the
  elementary charge $e$), are obtained in a Bader analysis
  \cite{Bader1985}; these charges are directly transferred to our
  classical model of the ClO$_4^-$ molecule.}
\label{tbl:perchlorate:xyzq}
\end{table}

\begin{table}[htbp]
  \centering
  \adjustbox{max width=\textwidth}{%
  \begin{tabular}{l c c c}
  \toprule
    & on top (OT) & side by side (SBS) & SBS on gold support \\
  \midrule
   PQP$^+ $                   &  1.0 & 0.9  &  0.9   \\
   \hspace{.5cm} N  ($M = 1$)        &  -1.1 & -1.1   &  -1.2\\
   \hspace{.5cm} C  ($M = 29$)    &  0.2 & 0.3  &  0.3  \\
   \hspace{.5cm} H  ($M = 18$)    &  1.9 & 1.7  &  1.8 \\
   ClO$_4^-$                    &  -1.0 & -0.9  &  -1.0   \\
   \hspace{.5cm} Cl ($M = 1$)    &  2.6 & 2.5  &  2.7   \\
   \hspace{.5cm} O ($M = 4$)   &  -3.6   & -3.5 & -3.7  \\
   Au (M = 72)  &   --- & ---  &  0.1  \\
  \bottomrule
  \end{tabular}
  }
  \caption{
    Charges obtained in Bader analysis for PQP$^+$, ClO$_4^-$
    ions in the gas phase and supported by the gold surface. The
    \textit{on top} (OT) and \textit{side by side} (SBS)
    configurations correspond to the configurations of single
    PQPClO$_4$ pairs as in Figs. 3 a) and b) of the main text,
    respectively.  The SBS configuration on gold is depicted in
    Fig.~\ref{fig:fit:BE for different gold layers}a).  The atom
    specific charges are summed values of all $M$ atoms of the same
    type.  }
  \label{tab:bader_charges}

\end{table} 

Table \ref{tab:bader_charges} compares the Bader
charges of single PQP$^+$ and ClO$_4^-$ pairs in the gas phase and
on the gold surface. This analysis reveals that the local charges
on the ions are very similar in both environments and practically
unaffected by the presence of the metal surface.

\FloatBarrier

\subsection{Image charges in a solvent}
\label{app:image_charges}

We may describe the electrostatic interaction between
a charge distribution and a metallic surface by the method of image
charges \cite{Jackson1999}.  The electric field inside a metal is
completely screened and therefore the field on the surface can only
be perpendicular to the metal surface.

Inside media the Maxwell equation for the displacement field
$\mathbf{D}({\bf r}, t)$ reads

\begin{equation}
  \nabla\cdot \mathbf{D} = -4\pi\epsilon_0 \rho
  \label{eq:maxwell:medium:D}
\end{equation}
where $\epsilon_0$ is the permittivity of the vacuum and $\rho({\bf
  r}, t)$ a charge density.  The relative permittivity $\epsilon_r$ of
the medium connects to the electric field $\mathbf{E}({\bf r}, t)$ to
$\mathbf{D}({\bf r}, t)$ via $\mathbf{D}({\bf r}, t) = \epsilon_0
\epsilon_r \mathbf{E}({\bf r}, t)$.
The Maxwell equation (\ref{eq:maxwell:medium:D}) then becomes

\begin{equation}
  \nabla\cdot \epsilon_r \mathbf{E}({\bf r}, t) = -4\pi\rho({\bf r}, t).
  \label{eq:maxwell:medium:E}
\end{equation}

A spatially constant permittivity $\epsilon_r$ thus leads to

\begin{equation}
  \nabla\cdot \mathbf{E}({\bf r}, t) = -4\pi\frac{\rho({\bf r}, t)}{\epsilon_r}
  \label{eq:maxwell:medium:E:scaled}
\end{equation}
which is the Maxwell equation in vacuum with a charge density scaled
by the relative permittivity.

A slab geometry with a planar, perfectly conductive lower
surface of infinite extend, separating the slab region from the
interior bulk region at $z=0$ can be modeled by image charges.
These are placed within the bulk according to
Eq.~(7) in the main text, i.e.  $\mathbf{r}_{i}
= (x_i, y_i, z_i)\rightarrow (x_i, y_i, -z_i) =
\mathbf{r}_{i^\prime}$ and $q_{i^\prime}\rightarrow -q_i$ and
fulfill metallic boundary conditions.
Including the constant electrostatic field $E_z$ introduced in
Eq.~(10) in the main text this
setting corresponds to the electrostatic potential
$\Phi(\mathbf{r})$ of the form

\begin{equation}
  \Phi(\mathbf{r})=\frac{1}{4\pi\epsilon}\sum_{i=1}^{n}q_i\left(
    \frac{1}{|\mathbf{r} -\mathbf{r}_i|}
  -\frac{1}{|\mathbf{r} -\mathbf{r}_{i^\prime}|}
  \right) + z\,E_z\,\Theta(z),
\end{equation}
where $\Theta(z)$ is the Heaviside Theta (or step) function which is 0
if $z\leq 0$ and 1 otherwise.  The potential vanishes at $z=0$ in the
entire $x, y$ plane, i.e.  $\Phi\left(\mathbf{r}_{xy}=(x, y,
z=0)\right) = 0$.  In fact, every term in the sum vanishes separately
as we have $|\mathbf{r}_{xy} -\mathbf{r}_i|^2 = (x-x_i)^2 + (y-y_i)^2
+ (\pm z_i)^2 = |\mathbf{r}_{xy} - \mathbf{r}_{i^\prime}|^2$, ensuring
metallic boundary conditions\cite{Jackson1999}.

\section{Identifications of self-assembly scenarios}
\label{app:self_assembly}

\subsection{Angle-axis gradient as calculated from the torque}
\label{app:torque}

The software package LAMMPS \cite{Plimpton1995} allows to evaluate
forces and torques of rigid molecules enclosed in a simulation box.
Since we are interested in the gradient of the potential energy with
respect to angle-axis vectors, $\MoleculeOrientation{}$, i.e.,
$\nabla_{\MoleculeOrientation{}}U = (\frac{\partial U}{\partial
  P_{1}}, \frac{\partial U}{\partial P_{2}}, \frac{\partial
  U}{\partial P_{3}})$, we present here the transformation which is
required to transform a three dimensional torque $\mathbf{T}=(T_x,
T_y, T_z)$ to an angle-axis gradient,
$\nabla_{\MoleculeOrientation{}}U$, or in a component-wise notation
$\frac{\partial U}{\partial P_{i}}=\partial_i U$, using Latin indices
$i=1,2,3$ for three dimensional vectors.

In LAMMPS orientations are expressed in terms of unit-quaternions via
the four dimensional vector $\mathbf{Q}^{(4)}$ (using Greek indices
$\nu = 0, 1, 2, 3$)

\begin{equation}
{\bf Q}^{(4)} =
(\cos{\frac{\theta}{2}}, \sin{\frac{\theta}{2}}\,\hat{\MoleculeOrientation{}})
= (Q_0, Q_1, Q_2, Q_3),
\label{eq:torque:q}
\end{equation}
where, as described in Subsection~\ref{app:axis_angle_framework} of
the S.I., $\theta=|\MoleculeOrientation{}|$ is the angle of rotation
around the axis
$\hat{\MoleculeOrientation{}}=\MoleculeOrientation{}/\theta$ and
$|\mathbf{Q}^{(4)}|=1$.

Following the documentation of LAMMPS \cite{Miller2002} the resulting
(four dimensional) torque vector, ${\bf T}^{(4)} = (0, T_x, T_y, T_z)
= (0, {\bf T})$ on a rigid body is specified via

\begin{equation}
  \mathbf{T}^{(4)}=-\frac{1}{2} \mathbb{S}^T_{4 \times 4} \nabla_{\mathbf{Q}^{(4)}}U +
  \mathbf{T}^{(4)}_{\mathrm{int}} .
  \label{eq:torque:4}
\end{equation}
The internal torque $\mathbf{T}^{(4)}_{\mathrm{int}}$, provided by
LAMMPS \cite{Plimpton1995}, ensures that $T^{(4)}=(0,T_x,T_y,T_z) =
(0, \mathbf T)$.

Henceforward the matrix index ``$4 \times4$'' indicates four-by-four
matrices.  In the above relation we have introduced the orthogonal
skew-matrix $\mathbb{S}_{4 \times 4}$, given by

\begin{equation}
  \mathbb{S}_{4 \times 4} =
  \begin{pmatrix}
     Q_0 & -Q_1 & -Q_2 & -Q_3 \\
     Q_1 &  Q_0 & -Q_3 &  Q_2 \\
     Q_2 &  Q_3 &  Q_0 & -Q_1  \\
     Q_3 & -Q_2 &  Q_1 &  Q_0 \\
  \end{pmatrix};
\end{equation}
further, $\nabla_{{\bf Q}^{(4)}}U$ is the gradient of the potential
energy with respect to the unit-quaternion vector ${\bf Q}^{(4)}$, or,
alternatively, in a component-wise notation $\frac{\partial
  U}{\partial Q_{\nu}}=\partial_\nu U$, $\nu=0,1,2,3$.

Since we are dealing with rigid bodies and explicitly avoid
intra-molecular interactions we can neglect
$\mathbf{T}^{(4)}_{\mathrm{int}}$ in Eq.~(\ref{eq:torque:4}); further,
$\mathbb{S}_{4 \times 4}$ is an orthogonal matrix, thus $\mathbb{S}_{4
  \times 4}\mathbb{S}_{4 \times 4}^T = \mathbb{S}_{4 \times
  4}^T\mathbb{S}_{4 \times 4} = \mathbb{I}_{4\times 4}$ where
$\mathbb{I}_{4\times 4}$ is the four dimensional unit matrix. Hence we
can rewrite Eq.~(\ref{eq:torque:4}) as

\begin{equation}
  \nabla_{{\mathbf Q}^{(4)}} U=-2\mathbb{S}_{4 \times 4}{\mathbf T}^{(4)},
  \label{eq:torque:grad_q}
\end{equation}
and further express $\partial_i U$ in terms of
Eq.~(\ref{eq:torque:grad_q}) using the chain rule
\begin{equation}
  \partial_i U = \frac{\partial U}{\partial P_{i}} = \frac{\partial Q_\nu}{\partial P_{i}}\frac{\partial U}{\partial Q_{\nu}} = (\mathbb{Q}_{3 \times 4})_{i \nu} \partial_{\nu} U ~~~~ i = 1, 2, 3 ~~~ {\rm and} ~~~ \nu = 0, 1, 2, 3 ;
  \label{eq:torque:grad_p}
\end{equation}
above we have used the Einstein summation convention.  For convenience
we have introduced the $3\times4$ matrix $\mathbb{Q}_{3 \times 4}$
with components $\mathbb{Q}_{i \nu} = \frac{\partial Q_\nu}{\partial
  P_{i}}$, or equivalently, $\mathbb{Q}_{3 \times 4} = (\frac{\partial
  Q_0}{\partial \mathbf{P}}, \frac{\partial \mathbf Q}{\partial
  \mathbf{P}})$, with components that -- using Eq.~(\ref{eq:torque:q})
-- can be written as

\begin{eqnarray}
  \frac{\partial Q_0}{\partial \mathbf{P}} & =
  & \left(-\frac{1}{2\theta} \sin{\frac{\theta}{2}} \right) \mathbf P
  \textnormal{, ~~~~~~~~~ and}
  \label{eq:torque:q_grad_p_0}
  \\
  \frac{\partial \mathbf{Q}}{\partial \mathbf{P}} &=&
    \frac{2}{\theta^2}\left(
      \cos{\frac{\theta}{2}}
     -\frac{1}{2\theta}\sin{\frac{\theta}{2}}
    \right)\,\mathbf{P}\cdot\mathbf{P}^T
    + \left(
      \theta^2 \sin{\frac{\theta}{2}}
    \right)\,\mathbb{I};
  \label{eq:torque:q_grad_p_123}
\end{eqnarray}
here $\mathbf Q=(Q_1, Q_2, Q_3)$, $\mathbf{P}\cdot\mathbf{P}^T =
\theta^2(\Skew{P}\Skew{P} + \mathbb{I})$, where $\mathbb{I}$ is again
the ${3\times3}$ identity matrix, the dot represents a dyadic product,
and $\Skew{P}$ is defined in Eq.~(\ref{eq:rotation:skew}).

With relation~(\ref{eq:torque:grad_q}) and using $\mathbb{Q}_{i\nu}$,
given by Eqs.~(\ref{eq:torque:q_grad_p_0}) and
(\ref{eq:torque:q_grad_p_123}), we can rewrite
Eq.~(\ref{eq:torque:grad_p}) as

\begin{equation}
  \nabla_{\mathbf P} U = -2\,\mathbb{Q}\,\mathbb{S}\,T^{(4)} = \mathbb{P}\,\mathbf T
\end{equation}
with the $3\times 3$ matrix

\begin{equation}
  \mathbb{P} = \frac{1}{\theta}\left[ (\cos{\theta} -
    1)\mathbb{I} +(\sin{\theta} -
    \theta)\,\Skew{P} \right] \Skew{P} -
  \mathbb{I},
\end{equation}
and with $\mathbf T=(T_x, T_y, T_z)$ being the torque in Cartesian
coordinates in the lab-frame.

\subsection{Order parameters}
\label{app:order_parameters}

In order to quantify the structural difference between configurations
identified via the optimization procedure we associate a feature
vector (i.e., a set of order parameters), $\mathbf f$, to every
configuration, $\Genome{}$ as defined by Eq.~(8) of the
main text. In this work we mainly rely on the so-called bond
orientational order parameters (to be denoted by $\Psi (\nu)$) defined
in Refs.~\citenum{Strandburg1984,Strandburg1988,Mazars2008}, which
provide information about the positional order of ordered structures,
and two variants of orientational order parameters (to be denoted by
$\alpha$ and $\beta$) put forward in Ref.~\citenum{Georgiou2014} which
correlate spatial and orientational degrees of freedom.

The evaluation of local order parameters strongly depends on the
method on how to identify neighbors: in this contribution we use the
well-defined method of Voronoi construction
\cite{LejeuneDirichlet1850,Voronoi1908}. Further, the above order
parameters are defined for two dimensional systems.  In our case of a
quasi-two dimensional geometry, with a molecular self-assembly in a
plane and with slightly stacked 3D structures, we use for the
calculation of the order parameters the projected coordinates of all
molecules to the $z=0$ plane.

All of these order parameters describe global properties (or
symmetries) of an ordered structure based on the local proximity of
its atomic or molecular entities.  To be more precise these parameters
can be expressed as a sum over local order parameters $\psi_i$,
calculated for all $N$ particles (or in our case molecules) in the
system, i.e.,

\begin{equation}
    \Psi(\nu) \sim \sum\limits_{i=1}^{N}\psi_i(\mathbf r_i, \mathbf r^{{\mathcal N}_i}; \mathbf u_i, \mathbf u^{{\mathcal N}_i};\nu) ;
\end{equation}
here the $\mathbf r_i$ and $\mathbf u_i$ specify the position and the
orientation of particle $i$, while $\mathbf r^{{\mathcal N}_i}$ and
$\mathbf u^{{\mathcal N}_i}$ are the set of positions and orientations
of the neighboring particles, respectively; ${\mathcal N}_i$ is the
number of neighbors of particle $i$; the role of $\nu$ will be
specified below.

To be more specific: bond orientational order parameters, $\Psi
(\nu)$, as defined in Refs.~\citenum{Strandburg1984,Strandburg1988}
and revisited in Ref.~\citenum{Mazars2008}, depend only on the
relative angle, $\phi_{ij}$ (with $\cos\phi_{ij}=\mathbf{\hat
  r}_{ij}\cdot\mathbf{\hat e}_\mathrm{ref}$), which is enclosed
between the bonds of a central particle, $i$, and the bonds to each of
its $\mathcal{N}_i$ neighbors

\begin{equation}
    \Psi(\nu)=\frac{1}{N}\sum\limits_{i=1}^N\left|\frac{1}{\mathcal{N}_i} \sum\limits_{j=1}^{\mathcal{N}_i}\exp[\imath\nu\phi_{ij}] \right|;
    \label{eq:bond_orientational_order_parameter}
  \end{equation}
here we introduce the vector $\mathbf{\hat r}_{ij} = (\mathbf{r}_{j} -
\mathbf{r}_{i})/|\mathbf{r}_{j} - \mathbf{r}_{i}|$, i.~e., the unit
vector between two neighbouring particles $i$ and $j$, the reference
axis $\mathbf{\hat e}_\mathrm{ref}$ (which is of unit length), and the
complex unity $\imath$, $\imath^2=-1$.  The orientational symmetry is
quantified by the (integer) variable $\nu$: the $\nu$-fold bond
orientational order parameter, $\Psi(\nu)$, assumes the value of one
if the angles between neighbors are multiples of $2\pi/\nu$ and
attains values close to zero for a disordered particle arrangement or
if the $\nu$-fold symmetry is not present.  However, the lattices we
are dealing with are never perfect and the number of nearest neighbors
can differ from the ideal value. These issues make the evaluation of
bond-orientational order parameters numerically unstable, even for
tiny deviations of the particle positions from the ideal configuration
\cite{MoritzAntlanger2015}. In an effort to guarantee, nevertheless,
numerical stability in the evaluations of the $\Psi(\nu)$, we use a
modification of the above defined bond orientational order parameter,
which was proposed in Ref.~\citenum{Mickel2013}: this modified
definition includes a weighting factor which is related to the polygon
side length, $l_{ij}$, that neighboring particles share

\begin{equation}
    \Psi(\nu)=\frac{1}{N}\sum\limits_{i=1}^N
    \left|
    \frac{1}{L_i}
    \sum\limits_{j=1}^{\mathcal{N}_i}l_{ij}\exp[\imath\nu\phi_{ij}]
    \right|;
    \label{eq:bond_orientational_order_parameter:weighted}
\end{equation}
with $L_i=\sum\limits_{j=1}^{\mathcal{N}_i} l_{ij}$; the polygon side
lengths, $l_{ij}$, are extracted from the Voronoi construction.

Since the molecules (and hence their interactions) are anisotropic, it
is useful to quantify their orientational order.  Similar to
Eq.~(\ref{eq:bond_orientational_order_parameter:weighted}) we can
quantify global {\it orientational} order including again the above
Voronoi nearest-neighbor construction in the following way:

\begin{equation}
    \beta=\frac{1}{2N}\sum\limits_{i=1}^N
      \frac{1}{L_i}
      \sum\limits_{j=1}^{\mathcal{N}_i}
        l_{ij} \left| \mathbf{\hat u}_i \cdot \mathbf{\hat u}_j\right| .
    \label{eq:orientational_order_parameter:beta}
\end{equation}
Finally, we can combine orientational order with positional degrees of
freedom, using the unit vector $\hat{\mathbf r}_{ij}$ between two
neighboring particles $i$ and $j$:

\begin{equation}
    \alpha=\frac{1}{2N}\sum\limits_{i=1}^N
      \frac{1}{L_i}
      \sum\limits_{j=1}^{\mathcal{N}_i}
        l_{ij} \left| (\mathbf{\hat u}_i \cdot \mathbf{\hat r}_{ij})^2 + (\mathbf{\hat u}_j \cdot \mathbf{\hat r}_{ij})^2\right|,
    \label{eq:orientational_order_parameter:alpha}
\end{equation}
suggesting again a modified version of the order parameters with the
side lengths of the Voronoi polygons, $l_{ij}$ \cite{Georgiou2014};
$\mathbf{\hat u_i}$ is here a unit-vector defining the orientation of
a particle in the lab-frame.  For a molecule, specified by the index
$\Molecule{I}$ we evaluate the orientation, $\hat{\mathbf
  u}_{\Molecule{I}}$, by rotating a reference vector, $\hat{\mathbf
  e}_{\mathrm{ref}}=\hat{\mathbf e}_x$, according to the current
angle-axis vector, $\MoleculeOrientation{I}$: $\hat{\mathbf
  u}_{\Molecule{I}} =
\mathbb{T}(\MoleculeOrientation{I})\cdot\hat{\mathbf
  e}_{\mathrm{ref}}$ (see Subsection~\ref{app:axis_angle_framework} of
the S.I. for details).

For our PQP$^+$ ClO$_4^-$ system we used a set of order
parameters to define the feature vector $\mathbf f$:

\begin{equation}
    \begin{split}
    \mathbf f=( &
    \Psi_{\mathrm{PQP}}(4), \Psi_{\mathrm{PQP}}(5), \Psi_{\mathrm{PQP}}(6),\\ &
    \Psi_{\mathrm{ClO}_4}(4), \Psi_{\mathrm{ClO}_4}(5), \Psi_{\mathrm{ClO}_4}(6),\\ &
    \Psi_{\mathrm{PQP|ClO}_4}(4), \Psi_{\mathrm{PQP|ClO}_4}(5), \Psi_{\mathrm{PQP|ClO}_4}(6),\\ &
    \beta_{\mathrm{PQP}}, \beta_{\mathrm{ClO}_4},\\ &
    \alpha_{\mathrm{PQP}}, \alpha_{\mathrm{ClO}_4}
    );
    \end{split}
    \label{eq:order_parameters}
\end{equation}
here $\Psi_{\mathrm{PQP}}(\nu)$ and $\Psi_{\mathrm{ClO}_4}(\nu)$
quantify the $\nu=4,5,6$-fold bond-orientational order parameters (see
Refs.~\citenum{Strandburg1984,Strandburg1988,Mazars2008}), defined by
Eq.~(\ref{eq:bond_orientational_order_parameter:weighted}),
considering only PQP$^+$ and ClO$_4^-$ molecules as neighbors.
$\Psi_{\mathrm{PQP|ClO}_4}(\nu)$ quantifies the $\nu=4,5,6$-fold
bond-orientational order parameters for all PQP$^+$ molecules while
considering only ClO$_4^-$ molecules as neighbors.
$\beta_{\mathrm{PQP}}$ and $\beta_{\mathrm{ClO}_4}$, defined by
Eq.~(\ref{eq:orientational_order_parameter:beta}), quantify the
orientational-correlation between neighboring PQP$^+$ and ClO$_4^-$
ions, respectively, while $\alpha_{\mathrm{PQP}}$ and
$\alpha_{\mathrm{ClO}_4}$, defined by
Eq.~(\ref{eq:orientational_order_parameter:alpha}), are in addition
sensitive to the respective spatial correlation between neighbours
(see Refs.~\citenum{Georgiou2014,MoritzAntlanger2015}).

\section{Results}
\subsection{General remarks and system parameters}
\label{app:results:remarks}

The following details provide an idea about the numerical costs of our
calculations: in order to obtain the ground state configuration for a
single state point (specified by a set of the system parameters
defined in Subsection~4.1 in the main text)
convergence of the full EA+LG ground-state search (based on the
evolutionary algorithm (EA) and the local, steepest gradient descent
procedure (LG) as specified in Section~3 in the
main text) we require at least one to two weeks on one node on the
Vienna Scientific Cluster (VSC3)
(\href{http://typo3.vsc.ac.at/systems/vsc-3/}{http://typo3.vsc.ac.at/systems/vsc-3/};
equipped -- per node -- with either two Intel Xeon E5-2650v2, 2.6 GHz,
eight core processors or two Intel Xeon E5-2660v2, 2.2 GHz, ten core
processors from the Ivy Bridge-EP family).  We typically used 16 to 20
asynchronous worker processes per evolutionary optimization.

\subsection{Clustering of results by similarity}
\label{app:clustering}

In Subsection~4.2.3 of the main
text (see, in particular
Fig.~6) we present
and discuss results which originate from an independent and separate
evolutionary algorithm analysis which focuses entirely on the
mobility of the perchlorate anions. This is done by fixing the unit
cell as well as the positions and orientations of the contained
PQP$^+$ molecules such that we can study the local minima in the
potential energy as a function of the degrees of freedom of the
ClO$_4^-$ molecules.  To be more specific these investigations were
performed for the structures depicted in
Fig.~4(a,e,c) of the main
text. Among the solutions identified by the evolutionary algorithm
we chose roughly 5000 configurations, all representing a local
energy minimum in this energy landscape, for which we evaluated the
set of order parameters, $\mathbf{f}$, specified by
Eq.(\ref{eq:order_parameters}) in
Subsection~\ref{app:order_parameters} of the SI. Here we extended
this set of order parameters for the problem at hand (i) with
additional bond orientational order parameters $\Psi_{\mathrm{ClO}_4}(\nu)$
and $\Psi_{\mathrm{PQP}|\mathrm{ClO}_4}(\nu)$ for $\nu=3,8$, and
$\Psi_{\mathrm{ClO}_4|\mathrm{PQP}}(\nu)$ for $\nu=3,4,5,6,8$,
defined analogously to
Eqs.~(\ref{eq:bond_orientational_order_parameter:weighted}
\ref{eq:order_parameters}), and (ii) with additional orientational
order parameters $\alpha_{\mathrm{ClO}_4|\mathrm{PQP}}$ and
$\beta_{\mathrm{ClO}_4|\mathrm{PQP}}$ as defined by
Eqs.~(\ref{eq:orientational_order_parameter:beta},
\ref{eq:orientational_order_parameter:alpha})
but considering the orientational order of perchlorate molecules
only with respect to PQP$^+$ neighbors and vice versa, by
$\alpha_{\mathrm{PQP}|\mathrm{ClO}_4}$ and
$\beta_{\mathrm{PQP}|\mathrm{ClO}_4}$.
Further, (iii) we used the minimum, mean, median and maximum
value of the $z$ coordinates of all ClO$_4^-$ ions and (iv) the
argument of the first peak in the radial distribution function
of the perchlorates and (v) of all molecules.

In order to identify configurations which are similar
in their structure among all these particle arrangements and to
further distinguish between different collections of similar
structures in these large sets of data we used unsupervised
clustering techniques; a very instructive review on such useful
tools and many other helpful machine learning applications in
physics or chemistry can be found in Ref.~\citenum{HIGHBIAS2019}.

To be more specific, we combine here the so-called
principal component analysis (PCA) \cite{Jolliffe2002} which reduces
the dimensionality of our structural data (or better order
parameters thereof) and a successive t-stochastic neighbor embedding
(t-SNE)\cite{Maaten2008} in order to map high-dimensional data
points to low-dimensional embedding coordinates (in two or three
dimensions), while preserving the local structure in the data
\cite{HIGHBIAS2019}.  With this tool at hand we aim at representing
high-dimensional data in two or three dimensions in order to unravel
hidden -- or hard to identify -- geometries (such as structural
similarities) within the data set.

Coming back to the discussion on the study of the
perchlorate molecules we narrowed down the collection of different
structures by admitting only such molecular arrangements whose
energy is located within a certain threshold interval, $\Delta E$,
above the energy of the related best structures, respectively.

The respective set of order parameters of those
structures is first scaled to unit-variance-- and zero-mean
coordinates considering all data-points. Subsequently, this scaled
set is subject to a PCA which is a linear transformation
of the $F$-dimensional data space (in our case of the order
parameters or features), $\mathbf{f}$, to a $L$-dimensional latent
space, $\mathbf{f}^\prime = \mathbf{f}\cdot\mathbb{L}$, using the
$F\times L$ projection matrix, $\mathbb{L}$; this is done with the
intention to identify leading singular values of the correlation
matrix of the data which represent directions in the data with large
variance, i.e. which contain the most relevant information
\cite{HIGHBIAS2019}. Thereby we reduced the number of our order
parameters from 32 to five for the results presented in
Fig.~6 of the main
text.

\begin{figure}[htbp]
\begin{center}
\includegraphics[width=0.7\textwidth, clip=True]{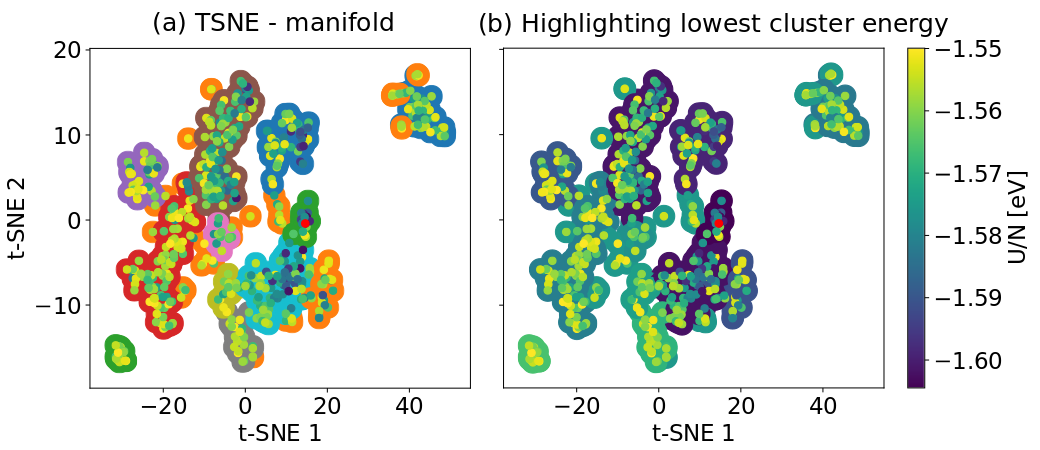}
\caption{(color online) Dimensional reduction of
    order parameters into a two dimensional manifold spanned by
    \textit{t-SNE 1} and \textit{t-SNE 2} using t-SNE
    clustering\cite{Maaten2008} for the results presented in
    Fig.~6 of the
    main text, in order to identify structurally different
    configurations.  (a): separated clusters represent structurally
    different configurations.  The energy of each configurations is
    color-coded (small dots) and given per PQP$^+$ -- ClO$_4^-$ pair,
    $N$.  Labels of clusters are highlighted using larger, uniformly
    colored dots in the background.  (b): same as left but the
    background-colors of each configuration indicates the minimal
    cluster energy.  The emphasized red dot indicates the structure
    with the lowest energy.  }
\label{fig:ea:results_perchlorate_stacked_clustering}
\end{center}
\end{figure}

These five leading principal components,
$\mathbf{f}^\prime$, which exhibit an explained variance above 5\%,
are then subject to a t-SNE analysis.  The idea behind such an
analysis is to identify neighbouring data-points (in our case in the
latent space after the PCA): this is done by matching the
probability distribution $p_{i|j}\propto \exp(-|\mathbf{f^\prime}_i
- \mathbf{f^\prime}_j|^2/(2\sigma_i^2))$, which quantifies the
likelihood that $\mathbf{f^\prime}_j$ is a neighbour of
$\mathbf{f^\prime}_i$ (where $\sigma_i$ is a bandwidth parameter and
is usually determined by fixing the local entropy $H(p_i)=-\sum_j
p_{j|i}\log_2{(p_{j|i})}$) with a similar probability distribution
$q_{ij}\propto (1+|\mathbf{t}_i-\mathbf{t}_j|^2)^{-1}$ by minimizing
the Kullback-Leibler divergence,
$D_{\mathrm{KL}}(p||q)=\sum_{ij}p_{ij}\log{(p_{ij}/q_{ij})}\rightarrow\min$,
between $q_{ij}$ and the symmetrized probability distributions
$p_{ij}=(p_{i|j}+p_{j|i})/(2N)$; the indices $i$ and $j$ run over the
number of data-points.  Neighbouring data points with small
separations, $|\mathbf{f^\prime}_i - \mathbf{f^\prime}_j|^2$ and
$|\mathbf{t}_i - \mathbf{t}_j|^2$, contribute most to $p_{ij}$ and
$q_{ij}$, respectively; however, in contrast to the $p_{ij}$ latent
space data-points, $\mathbf{t}_i$ and $\mathbf{t}_j$, are explicitly
encouraged to repel each other if they are far apart by the tails of
$q_{ij}$.  This approach can be interpreted as equilibrating the
effects of attractive forces acting on neighbouring data-points
(being grouped within a cluster) and repulsive forces between whole
clusters (separating different groups).  Hence, it is easier to
visually separate clusters in a few t-SNE dimensions than in a
higher dimensional PCA latent space \cite{HIGHBIAS2019}; the results
of a two-dimensional t-SNE analysis of our problem at hand can also
be observed in
Fig.~\ref{fig:ea:results_perchlorate_stacked_clustering}.  In
addition, we performed a DBSCAN \cite{Ester1996}, i.e., a
density-based clustering algorithm which automatically labels the
different clusters; these labels are highlighted in the left panel
of this figure by color coding.

The minimum energy of each cluster of structures is
indicated in the right panel of this figure by a color code; the
results clearly indicate, that several clusters are separated by
only minute energy differences.  Four of these configurations,
separated by an energy difference of $38$~meV per molecule pair, are
shown in Fig.~6 of
the main text.

\bibliography{main}